\documentclass[twocolumn,aps,prb,longbibliography]{revtex4-1}
\usepackage{amsmath,amssymb} 
\usepackage{amsmath} 
\usepackage{amsfonts} 
\usepackage{bm} 
\usepackage[colorlinks=true,citecolor=blue,linkcolor=blue]{hyperref}
\usepackage[pdftex]{graphicx} 
\usepackage{braket} 

\newcommand{\ch}[1]{{\color{black} #1}}

\newcommand{\revs}[1]{{\color{black} #1}}
\newcommand{\revt}[1]{{\color{black} #1}}

\begin{document}

\title{Cluster-projected matrix product state: 
framework for engineering exact quantum many-body ground states in one and two dimensions}
\author{Hidehiro Saito}
\email{saito-hidehiro722@g.ecc.u-tokyo.ac.jp}
\author{Chisa Hotta}
\email{chisa@phys.c.u-tokyo.ac.jp}
\affiliation{Department of Basic Science, University of Tokyo, Meguro-ku, Tokyo 153-8902, Japan}
\date{\today}

\begin{abstract}
We propose a framework to design concurrently a frustration-free quantum many-body Hamiltonian and 
its numerically exact ground states on a sufficiently large finite-size cluster 
in one and two dimensions using an elementary matrix product state (MPS) representation. 
Our approach strategically chooses a local cluster Hamiltonian, which is arranged to overlap 
with neighboring clusters on a designed lattice. 
The frustration-free Hamiltonian is given as the sum of the cluster Hamiltonians 
by ensuring that there exists a state that has its local submanifolds 
as the lowest-energy eigenstate of every cluster. 
The key to find such a solution is a systematic protocol, 
which projects out excited states on every cluster using MPS and effectively entangles the cluster states. 
The protocol offers several advantages, 
including the ability to achieve exact many-body ground-state solutions at nearly equal cost 
in one and two dimensions, 
those belonging to gapless or long-range entangled classes of ground states, 
flexibility in designing Hamiltonians unbiasedly across various forms of models, 
and numerically feasible validation through energy calculations. 
Our protocol offers the exact ground state for general frustration-free Hamiltonian, 
and enables the exploration of exact phase boundaries and the analysis of even a spatially nonuniform random system, 
providing platforms for quantum simulations and benchmarks. 
\end{abstract}
\maketitle
\section{Introduction}
Understanding the internal structure of quantum many-body states 
provides insights into emergent material phases, which can give certain connectivity to quantum information processing. 
For instance, the quantum ${\mathbb Z}_2$ spin liquid phase\cite{Savary2017} 
was found to be the exact solutions of the Kitaev model and toric codes
\cite{Kitaev2006,Kitaev2003}, and related topological orders and the braiding statistics were recently demonstrated 
in a quantum simulators\cite{Satzinger2021,Semeghini2021}. 
Methodologically, however, finding such intriguingly entangled quantum states relies much on a matter of luck, 
as the clarification of highly entangled states is often hindered by 
the exponential growth of complexity with system size\cite{Calabrese2004}. 
Traditional numerical techniques like exact diagonalization (ED) 
give valuable insight as they provide us with the exact description of the states, 
but their scalability and computational cost limit their applicability in large-size regimes. 
\par
If the exact wave functions are available, 
they offer a playground for theorists to find new concepts, 
such as Laughlin wave function\cite{Laughlin1983} led to topological orders in quantum spin liquids,  
and the Affleck-Kennedy-Lieb-Tasaki (AKLT) states\cite{Affleck1987} 
have proved useful in discovering the symmetry-protected topological (SPT) phases\cite{Pollmann2010}. 
However, a class of so-called exact solutions are typically given in the analytical tractable or mathematically 
integrable form, and are very limited. 
\par
With these backgrounds, 
there is a demand to find an exact form of quantum many-body states more systematically and practically 
using a concrete numerical representation based on particular basis sets. 
Such numerically exact form is often very useful, 
as the first detection of quantum phase transitions in a quantum simulator was 
done using the exact solution of the ZXZ model that traverses the SPT and transverse 
Ising product phases\cite{Smith2022}. 
\par
In search of such an exact representation, 
our work particularly focuses on a class of Hamiltonian called ``frustration-free", meaning the ground state is the simultaneous ground state of all local Hamiltonians. 
The question of whether a given Hamiltonian is frustration-free is a quantum $k$-SAT problem (see Appendix \ref{app:k-qsat}) 
known as QMA$_1$-complete\cite{Kitaev2002,Bravyi2006,Gosset2016,Aldi2020}, namely, a class of problems that the solution 
cannot be verified by the classical polynomial-time algorithm. 
We then propose an algorithm that not only gives its answer at the practical level, 
but enable the systematic derivation of {\it the numerically exact ground state in a large-size lattice} 
based on the matrix product state (MPS) representation. 
The MPS\cite{Fannes1992,Perezgarcia2007} is the simplest and the most efficient tool 
among the tensor network techniques\cite{Ors2019} like 
infinite projected entangled pair states (iPEPS)\cite{Corboz2016} and
multiscale entanglement renormalization ansatz (MERA)\cite{Vidal2007}. 
The MPS protocol is developed based on density matrix renormalization group (DMRG) scheme\cite{White1992,Verstraete2023}, 
both serving as the best variational ansatz of one-dimensional (1D) quantum many-body states. 
Yet, MPS and PEPS face the area-law bound of entanglement entropy (EE), 
and the previous algorithm optimizing the local density matrix 
have so far been suitable for a gapped state or at most a critical state with logarithmic EE.  
\par
Our approach hinges on the non-TI form of MPS and obtains the wave function of the lattice 
size and bond dimensions up to which we can store the entanglement and numbers of degeneracy, 
tractable as the exact ground state. 
By selecting a local cluster Hamiltonian and its lowest energy eigenstates, 
we strategically design a lattice where clusters share sites with their neighbors 
to have a bulk Hamiltonian as the sum of cluster Hamiltonians. 
In building an exact ground state, we discard the previously established optimization algorithm of 
the variational ansatz, 
and instead, develop an algorithm that systematically applies a projector 
to fix the tensor elements one by one to properly entangle the cluster states. 
Most importantly, this algorithm is applied to any given candidate model, 
where we can immediately judge whether it is frustration-free, 
and if yes, simultaneously obtain the exact MPS ground state. 
We see shortly that this protocol practically works well as 
it can handle the exact ground state and even an excited state of the long-range-entangled topological states 
in two dimension (2D) of a size comparable to DMRG, far beyond what ED could attain. 
While inversely, it is known that for any given injective TI-MPS state, 
one can always obtain a frustration-free parent Hamiltonian\cite{Perezgarcia2007,Cirac2021}, 
and its ground state is proved to be unique and has a finite gap\cite{Perezgarcia2007}. 
The gapless state is not afforded possibly because of the restriction imposed to 
directly accesses the infinite-size state. 
Our pragmatic method is a spiritually different approach that safely describes such gapless classes of states 
by restricting the size of the lattice.  
\par
The paper is organized as follows. 
In \S.\ref{sec:clusterent}, we first highlight the core of the method of 
constructing a candidate Hamiltonian and verify whether it is frustration-free or not, 
and show demonstration on how the exact MPS eigenstate is obtained 
using the toric code model. 
Those who are interested in the details of calculating the MPS solution 
shall go to \S.\ref{sec:exactmps} where we present the algoritm using the example on the diamond lattice. 
The applications of the MPS protocol to the 2D model is shown in \S.\ref{sec:2d}, 
where one finds the quantitative information on which kind of lattices or models can have 
the solutions to what extent by tuning the parameters. 
In \S.\ref{sec:triangle} we further show how to find the exactly solved frustration-free model across the parameter 
space, where we demonstrate that different models have the same classes of exact solutions on the zigzag spin-1/2 chain. The details of the protocol and the other versions of algorithms off the main contexts are given in Appendices. 
Throughout this paper, we provide new examples on top of the well know frustration-free models (Table I) 
that are found using our protocol in \S.\ref{sec:2d}, \ref{sec:triangle}, and Appendix \ref{app:tri-mps}. 

\begin{figure*}
    \centering
    \includegraphics[width=18cm]{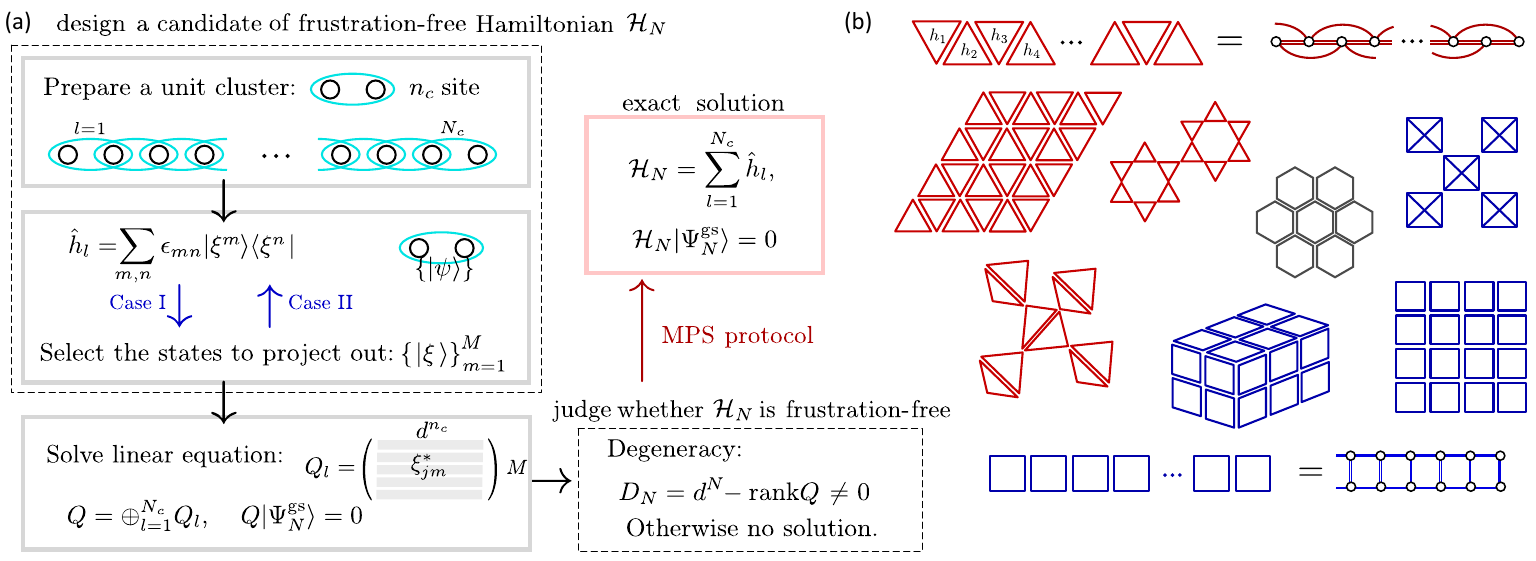}
    \caption{
\revt{(a) Flowchart, starting from designing a candidate of frustration-free Hamiltonian, 
judge whether ${\cal H}_N$ is frustration-free or not by the rank of $Q$, 
and if yes, construct an exact ground state solution using MPS protocol. } 
In designing, Case I first determines $\hat h_l$ and obtain its excited states, $|\xi\rangle$, 
and Case II is vice versa. 
(b) Illustration of typical lattice models constructed as a sum of unit clusters: 
zigzag ladder, ladder, triangular, kagome, pyrochlore, honeycomb, cube, and square lattices. 
In general, the corner-shared lattices in two and three dimensions are difficult to treat 
because of the small number of constraints $M$. 
}
    \label{f1}
\end{figure*}
\section{Designing a frustration-free Hamiltonian and exact solutions}
\label{sec:clusterent}
\subsection{Essence of the framework}
\label{sec:clusterent-1}
We first illustrate the essence of the present framework by 
using an AKLT\cite{Affleck1987} state, 
although most of the states we discuss are far more complex and are not written down in a 
simple analytical form like AKLT. 
Let us consider a unit consisting of $n_c$ sites, where each site carries $d$ degrees of freedom. 
Assembling these units by sharing their $n_\cap (<n_c)$ sites with its neighbors, 
we obtain a periodic lattice as shown in Fig.~\ref{f1}(a). 
\par
Our goal is to construct a quantum state on 
{\it a finite but sufficiently large $N$ site lattice}, $|\Psi_N^{\rm gs}\rangle$, 
that serves as an exact ground state of a Hamiltonian written as 
a sum of positive-definite operator $\hat h_l$ acting on the $l$-th cluster: 
\begin{equation}
{\cal H}_N=\sum_{l=1}^{N_c} \hat h_l. 
\label{eq:ham}
\end{equation}
Whenever we project $|\Psi_N^{\rm gs}\rangle$ onto 
any of the unit clusters by integrating out the $N-n_c$ part, 
it should consist of a specified manifold of states $\{|\psi\rangle\}$ on the $l$-th cluster, 
satisfying $\hat h_l |\psi\rangle=0$. 
If such $|\Psi_N^{\rm gs}\rangle$ is obtained, 
Eq.(\ref{eq:ham}) will be a ``frustration-free" Hamiltonian 
because the ground state energy is the sum of the lowest eigenvalues (zero) of $\hat h_l$. 
Here, the local Hibert space of a unit cluster has dimension $d^{n_c}$, 
which are classified by this penalty term $\hat h_l$ into two groups 
$\{|\psi\rangle\}$ and $\{|\xi\rangle\}$ of dimension $D_g$ and $M$, respectively, with $D_g+M=d^{n_c}$. 
\par
The AKLT model serves as the most elementary example. 
Let the spin-1 chain be regarded as a site-shared spin-1 dimers with $n_c=2$ and $d=3$, 
where we need $N_c=N$ dimers for the periodic boundary condition (PBC) and $N_c=N-1$ for 
the cluster open boundary condition(C-OBC). 
The cluster Hilbert space of dimension 9 is spanned by four $S=0,1$ states 
as $\{|\psi^{m}\rangle\}_{m=1}^4$ and five $S=2$ states as $\{|\xi^{m}\rangle\}_{m=1}^5$, 
where we want to project out the latter. 
Using a projection operator ${\cal P}_2$ onto $S=2$, we obtain a local penalty Hamiltonian,
\begin{equation}
\hat h^{\rm AKLT}_l= \sum_{m=1}^5 |\xi^{m}\rangle \langle \xi^{m}|
={\cal P}_2 (\bm S_l+\bm S_{l+1}). 
\end{equation}
By rewriting it using the spin-1 operator $\bm S_i$ and taking a sum over units, 
we reach the AKLT Hamiltonian: 
\begin{equation}
{\cal H}_N^{\rm AKLT}=\sum_{i=1}^{N} \frac{1}{2}(\bm S_i\cdot\bm S_{i+1})+\frac{1}{6}(\bm S_i\cdot\bm S_{i+1})^2+\frac{1}{3}. 
\end{equation}
The exact ground state $|\Psi_N^{\rm gs}\rangle$ that satisfies 
${\cal H}_N^{\rm AKLT} |\Psi_N^{\rm gs}\rangle=0$ is known as AKLT state. 
\par
More generally, one can consider any $n_c$-site unit, making it share a site, 
an edge or a plane with its adjacent units, 
and design a lattice model and its exact ground state. 
Suppose that among $d^{n_c}$ states, the manifold, $\{|\xi^{m}\rangle\}_{m=1}^{M}$, are designed to be projected out 
from the ground state. 
These states are orthogonal to the rest of the states $\{|\psi^{m'}\rangle\}_{m'=1}^{D_g}$ 
that constitute the ground state as, 
$\langle \xi^{m}|\psi^{m'}\rangle=0$. 
We set $\hat h_l$ as a local penalty Hamiltonian written in the form, 
\begin{equation}
\hat h_l= \sum_{m,n=1}^{M} \epsilon_{m n}|\xi^{m}\rangle\langle\xi^{n}|, 
\label{eq:localham}
\end{equation}
where the $M\times M$ matrix $\epsilon_{mn}$ should be positive definite 
to have $|\xi^{m}\rangle$ as excited states. 
\par
However, determining $\hat h_l$ and $\{|\xi^{m}\rangle\}$ 
does not guarantee that we can obtain an exact eigenstate of ${\cal H}_N$. 
Unlike the exact AKLT state known {\it a priori}, we need to 
derive the actual form of $|\Psi_N^{\rm gs}\rangle$ that satisfies 
\begin{equation}
{\cal H}_N |\Psi_N^{\rm gs}\rangle=0. 
\label{eq:gscond}
\end{equation}
Because one can always perform a Schmidt decomposition of any finite-size wave function 
into $l$-th cluster and the rest of the system of size $N-n_c$, 
whenever Eq.(\ref{eq:gscond}) holds, it means no nonzero Schmidt value  
for $\{|\xi^m\rangle \}$, in which case the ground state is expressed as 
\begin{equation}
|\Psi_N^{\rm gs}\rangle= \sum_{m=1}^{D_g} \lambda_m |\psi^{m}_l\rangle |\Phi_{\bar l}^{m}\rangle, 
\label{eq:gscond2}
\end{equation}
where $\{|\Phi_{\bar l}^{m}\rangle\}$ is the Schmidt state on the rest of the system. 
This holds for both PBC and C-OBC. 
Operating $\hat h_l$ immediately gives Eq.(\ref{eq:gscond}). 
However, entangling $\{|\psi^{m}\rangle\}$ with those of their neighbors 
to make it fulfill Eq.(\ref{eq:gscond2}) is not a promising task, 
and there can often be no solution that satisfies Eq.(\ref{eq:gscond}), 
in which case ${\cal H}_N$ is no longer called ``frustration-free". 
The core of our paper is the protocol given in Section \ref{sec:lineareq}, 
that judges whether ${\cal H}_N$ is frustration-free or not, and if yes, 
systematically provide $\{|\Psi_{N}^{\rm gs}\rangle\}$ satisfying Eq.(\ref{eq:gscond}) 
or equivalently (\ref{eq:gscond2}). 
The protocol is pinned down to a more pragmatic one that obtains a non-TI exact MPS 
in Sec.\ref{sec:exactmps}. 
\par 
In Fig~\ref{f1}(a) we show two strategies in choosing $\hat h_l$ and $\{\xi^m\}$.  
In Case I, $\hat h_l$ and ${\cal H}_N$ are given. 
We apply the protocol in \S.\ref{sec:lineareq} to check 
whether there is a solution of $|\Psi_N^{\rm gs}\rangle$ that satisfies Eq.(\ref{eq:gscond}), 
and if yes (i.e. $D_N\ne 0$ in Eq.(\ref{eq:dn})), 
${\cal H}_N$ is frustration-free, and we obtain $|\Psi_N^{\rm gs}\rangle$. 
\par
Case II applies when we want to search for an unknown Hamiltonian (see \S.\ref{sec:triangle} for demonstration). 
We vary the choice of cluster states $\{|\xi^{m}\rangle\}_{m=1}^{M}$ 
as well as how to overlap the clusters $n_\cap$. 
For each choice, $\hat h_l$ and ${\cal H}_N$ are tested, 
so as to find the choice that satisfies Eq.(\ref{eq:gscond}). 

\begin{table*}
\caption{
Previously established frustration-free models. 
The nature of the ground state (g.s.), whether it is gapped or gapless, the number of degeneracies for PBC/OBC (P/O), and the spatial dimensions are shown. }
\begin{tabular}{llccccccc}
\hline 
\rule{3mm}{0mm} &model \rule{25mm}{0mm} & g.s. & gap \rule{2mm}{0mm}
& degeneracy \rule{2mm}{0mm} & BC \rule{2mm}{0mm} & dim. \rule{2mm}{0mm}  &ref.  \rule{2mm}{0mm} \\
\hline
\hline
&AKLT chain &  SPT     & gapped  & 1/4  & P/O &1D (2D)   & [\onlinecite{Affleck1987,denNijs1989,Kennedy1992, Kennedy1992-2,Li2023}] \\
&Majumdar-Ghosh chain & product state &  gapped & 2 & P &1D  & [\onlinecite{Majumdar1969}, \onlinecite{Majumdar1969-2}] \\
&PXP-like chain &  liquid    &  gapped  & 1/4  & P/O &1D  & [\onlinecite{Mark2020}, \onlinecite{Lesanovsky2012}]\\
&Motzkin chain&  Motzkin walk  & gapless & 1 & P &1D  & [\onlinecite{Alexander2021}, \onlinecite{Bravyi2012}] \\
&Fredkin chain&  Dyck walk  & gapless & 1 &P & 1D  & [\onlinecite{Salberger2017}, \onlinecite{Salberger2017-2}] \\
&Zigzag XXZ chain &  anyon BEC    &  gapless  &  $O(N^2)$ & O &1D  & [\onlinecite{Batista2009}, \onlinecite{Batista2012}]\\
& Three-coloring problem & product state  & gapless  & many & any  & 2D (1D) & [\onlinecite{Changlani2019}, \onlinecite{Palle2021}] \\
& Kitaev's toric code& $\mathbb{Z}_2$ spin liquid  & gapped  & 4 & P   & 2D (3D)  & [\onlinecite{Kitaev2003}, \onlinecite{Verstraete2006}] \\
&Rokhsar-Kivelson point &  short ranged RVB \rule{2mm}{0mm}  & gapped & 1 & P  & 2D & [\onlinecite{Rokshar1988}, \onlinecite{Moessner2001}]  \\
\hline
\end{tabular}
\label{tab2}
\end{table*}
\subsection{Protocol for constructing exact eigenstate}
\label{sec:lineareq}
We start from an elementary method that straightforwardly imposes the condition Eq.(\ref{eq:gscond2}) 
(equivalent to Eq.(\ref{eq:gscond})) 
for all $l=1,\cdots, N_c$. 
These conditions reduce to a set of linear equations, 
and their solutions, if present, are the exact ground states $|\Psi_N^{\rm gs}\rangle$. 
Importantly, whether the solution exists or not can be easily judged. 
\par
Suppose that we decide to project out $M$ different states
on the $l$-th cluster, represented using the normalized and orthogonal set of 
$d^{n_c}$ basis $\{|x\rangle\}$ as 
\begin{equation}
|\xi^{m}\rangle= \sum_{x=0}^{d^{n_c}-1} \tilde \xi^{m}_{x} |x\rangle. 
\label{eq:projctedst}
\end{equation}
with $\tilde \xi_{x}^m\in {\mathbb C}$. 
The local Hilbert space is classified as 
$\big\{ \{|\psi^{m}\rangle\}_{m=1}^{D_g}, \{|\xi^{m'}\rangle\}_{m'=1}^{M} \big\}$. 
\par
For the system size $N$ consisting of $N_c$ cluster, 
the Hilbert space dimension is $d^N$, 
and its subspace with $l$-th cluster state being restricted to $\{|\psi^{m}_l\rangle\}$ is defined as 
\begin{equation}
V_l=\big\{ \{|\psi_l^{m}\rangle\} \otimes |\Psi_{\bar{l}}^{j}\rangle\big\}
_{m=1,\cdots,D_g; \;\; j=0,\cdots,d^{N-n_c}-1}, 
\label{eq:vl}
\end{equation}
where $\{ |\Psi_{\bar{l}}^{j}\rangle \}$ is the space spanned by $N-n_c$ sites 
that do not belong to the $l$-th cluster. 
\par
The ground states belong to the subspace $V_{\rm gs}\equiv \cap_{l=1}^{N_c} V_l$. 
Projection to it means imposing the following conditions on the whole Hilbert space: 
we prepare a $M\times d^{n_c}$ matrix using the coefficients of Eq.(\ref{eq:projctedst}), 
$Q_l=( \tilde \xi^{m*}_{x} )$, and the conditions are described by a set of linear equations 
represented by a matrix $Q$ with $d^{N-n_c} M N_c$ rows and $d^N$ columns as 
\begin{align}
 & Q |\Psi_N^{\rm gs}\rangle =0,  \notag\\
 & Q =\sum_{l=1}^{N_c} I^{d^{n_l}} \otimes Q_l\otimes I^{d^{N-n_c-n_l}}, 
   \label{eq:linear}
\end{align}
where $n_l$ is the number of sites on the left hand side of the $l$ th cluster. 
Solving this linear equation numerically gives the exact ground state 
when ${\rm dim} \cap_{l=1}^{N_c} V_l \ne 0$, 
and the degeneracy of the ground state is 
\begin{equation}
D_N={\rm dim} \cap_{l=1}^{N_{c}} V_l = d^N-{\rm rank} Q. 
\label{eq:dn}
\end{equation}
\revs{
Namely, by obtaining $D_N$, one could generally answer the quantum $k$-SAT question on whether ${\cal H}_N$ is frustration-free. 
}
Here, we mean by {\it ``exact"}, not the analytical tractability (formula) of the wave function. 
The numerical redundancy is natural because 
the wave functions leave the choice of phases or gauges, 
and also the choice of the orthogonal sets when $D_N\ge 2$. 
Whereas, the exactness is guaranteed by machine epsilon as local bases included are distinct,  
which makes the results numerically tractable. 
\par
Because the dimensions of $Q$ increase exponentially, 
there is an upper bound of $N$ that can be computed, 
which is comparable to the ED scheme. 
This cost can be reduced by subsequently linking one cluster to another and 
also by using the symmetries
\footnote{
\revt{
We first prepare a single unit $N_c=1$, 
imposing a condition Eq.(\ref{eq:projctedst}) 
to obtain a set of $\{\psi^m\}$ of dimension $D_g$. 
Then, add one unit cluster from another, where each time 
the condition to project the $d^{n}$ dimensional Hilbert space 
to $D_n$ solutions and the extra $M$ condition to project the states spanned on an added unit 
gives the number of rows of $Q$ as $D_n M$, which is reduced significantly from $d^{N-n_c}M N_c$. 
The solutions can be classified by the symmetry of ${\cal H}_N$ represented by the operator ${\cal R}$. 
When the Hilbert space of the $n$ site system is divided into $r_n$-sectors by this symmetry, 
$\{R_j\}_{j=1}^{r_n}$, one can project the states onto these sectors at each process of adding a unit cluster, 
and obtain the eigenstate of both ${\cal R}$ and ${\cal H}_N$. 
We show the example in \S.\ref{sec:udrvb}. 
}}. 
\par
However, in practice, we do not need to have large $N$ to judge whether the exact solution exists. 
The main usage of the protocol here is {\it to find a proper choice of $\{|\xi^{m}\rangle\}$ for a given cluster 
that yields a set of bulk Hamiltonian and the exact ground state. } 
Once we confirm that there is a solution, 
we shall shift to the method in Section \ref{sec:exactmps}, 
where the exact MPS tensors representing $|\Psi_N^{\rm gs}\rangle$ 
are determined similarly to Eq.(\ref{eq:linear}), 
that reaches far larger $N$ by formally ``compressing the information" without sacrificing the exactness. 
\par
Figure~\ref{f1}(b) shows examples of the choices of unit clusters and the lattices. 
The edge-shared lattices can be constructed both by the edge- and corner-sharing of clusters. 
However, the corner-shared units are not favorable for the present protocol in 2D as we see in \S.\ref{sec:2d} 
because $M$ is relatively small and the number of solutions increases exponentially. 
\par
In Table I we list a series of frusration-free models whose ground states are established. 
\revt{
They range from gapped 1D to gapless 2D states, 
highlighting that the well-known exactly solved models 
are concentrated on this class
(Here, the major class of exact integrable solutions are not frustration-free and are out of scope), 
while their solutions rely on model-specific languages. 
Some state rely on conserved quantum numbers\cite{Affleck1987, Lesanovsky2012,Kitaev2003}
e.g. the one on the zigzag chain is classified by the number of anyons\cite{Batista2012}. 
The MPS solutions are found in the Fredkin chain\cite{Salberger2017} and PXP-like chain\cite{Lesanovsky2012} 
and the MERA description in the Motzkin chain\cite{Alexander2021}. 
These forms are obtained by converting the patterns of ``walkers" or the 
preobtained exact analytical solutions to tensors, which are also not easy to identify. 
The present method offers numerically exact solutions for all the models in Table I, 
e.g. see \S.\ref{sec:aklttoric} and \S.\ref{sec:2d}), and for other unknown cases
as we see in \S.\ref{sec:2d} and \ref{sec:triangle}. 
}
\begin{figure*}
    \centering
    \includegraphics[width=18cm]{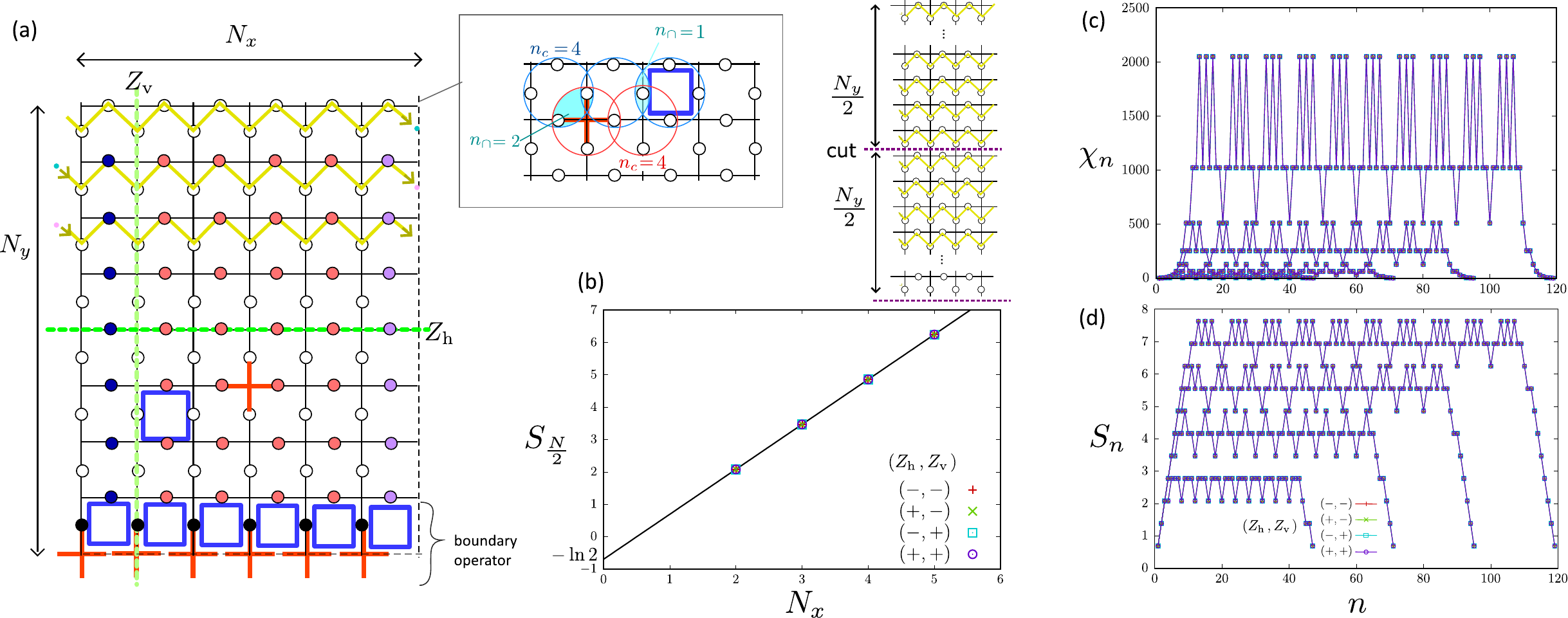}
    \caption{(a) Toric code in finite $N_x\times N_y$ size plaquettes including $N=2N_xN_y$ sites at the centers of the links. 
     The red crosses and blue plaquettes indicate the projector $A_v$ and $B_p$ and the two green lines 
     show the loop operators, $Z_{{\rm h}}$ and $Z_{{\rm v}}$. We color the sites depending on the types of projections 
     imposed to determine their MPS tensor in constructing the MPS along the yellow 1D paths (partially shown in the upper part). 
    (b) Entanglement entropy (EE) of the obtained exact MPS on four topological sectors 
        for the division of the system into half, $S_{N/2}$, given as a function of circumferences 
        $N_x$ at $N_y=12$. 
    (c) Bond dimension $\chi_n$ and (d) EE, $S_n$, for the bipartition into the first $n$ site and the rest $N-n$, 
    for four different topological sectors with \revs{$N_x=2,3,4,5$ ,$N_y=12$, namely up to $N=120$}. }
    \label{faps3}
\end{figure*}
\subsection{Revisiting AKLT and toric codes}
\label{sec:aklttoric}
Before explaining the technical details, 
we briefly show the applications to the AKLT and toric code to concretely get the idea. 
\par
For AKLT, we span a Hilbert space of spin-1 dimer as 
$\{|S^z_l,S^z_{l+1}\rangle\}=(|1,1\rangle,|1,0\rangle,|0,1\rangle,|1,-1\rangle,|0,0\rangle,\cdots,|-1,-1\rangle)$ 
and classify them into $S=0,1$ and $S=2$ manifolds. 
The projection matrix $Q_l$ consists of Clebsch-Gordan coefficients of $S=2$ states,  
\begin{equation}
Q_l=\left(\begin{array}{ccccccccc}
1 &0 & \cdots &&&&&& 0 \\
0 &\frac{1}{\sqrt{2}}& \frac{1}{\sqrt{2}} & 0 & \cdots &&&& 0 \\
0 &0&0& \frac{1}{\sqrt{6}} & \frac{2}{\sqrt{6}}&\frac{1}{\sqrt{6}} & 0 &0& 0\\
0 &\cdots&&&&0&\frac{1}{\sqrt{2}}& \frac{1}{\sqrt{2}} & 0 \\
0& \cdots &&&&&&0 & 1
\end{array}\right). 
\end{equation}
On the other hand, the MPS representation of the AKLT state for PBC is known as 
\begin{eqnarray}
&& |\Psi_N^{\rm AKLT}\rangle = \sum_{\{i_n\}} 
{\rm Tr} \big( \prod_{n=1}^N A^{i_n} \big) | i_1, i_2,\cdots, i_N\rangle, \nonumber \\
&&A^{\pm 1}=\pm \frac{2}{\sqrt{3}} \sigma^{\pm} ,\;\; A^{0}=-\frac{1}{\sqrt{3}} \sigma^z, 
\end{eqnarray}
where $i_n=1,0,-1$ denotes $S_n^z$ on site $n$ and $\sigma$ is the Pauli matrix. 
The operation, $Q \,|\Psi_N^{\rm AKLT}\rangle =0$, in Eq.(\ref{eq:linear}) corresponds to having 
\begin{eqnarray}
&&  A^1A^1=0 \nonumber \\
&&  A^1A^0 +A^0A^1=0  \nonumber \\
&&  A^1A^{-1} +2A^0A^0 +A^{-1}A^1=0  \nonumber \\
&&  A^{-1}A^{-1}=0, 
\label{eq:akltmps}
\end{eqnarray}
which is confirmed straightforwardly. 
We can also check that the OBC solution has a four-fold degeneracy 
that explains the number of edge states. 
\par
%
The toric code has the $\mathbb{Z}_2$ quantum spin liquid ground state\cite{Kitaev2003}. 
The Hamiltonian consists of vertex ($A_v$) and plaquette ($B_p$) operators as 
\begin{align}
 \mathcal{H}_{{\rm toric}} &= -\sum_v A_v -\sum_p B_p, \notag \\
& A_v = \prod_{i\in v}\sigma^x_i, \quad B_p = \prod_{i\in p}\sigma^z_i. 
\label{eq:torham}
\end{align}
All $A_v$'s and $B_p$'s commute and have eigenvalues $\pm 1$, 
so that the ground state will be the simultaneous $+1$ eigenstate of all the operators. 
which makes Eq.(\ref{eq:torham}) frustration-free. 
In applying our MPS protocol(see \S.\ref{sec:exactmps} and \ref{sec:2d}), 
we prepare two species of clusters both with $n_c=4$, given in blue and red circles 
in the inset of Fig.~\ref{faps3}(a), which have overlap $n_\cap=1$ and 2 
between the same and different species, respectively. 
The local penalty Hamiltonians for the two clusters are $\hat h_l=A_v$, $B_p$, and 
$Q_l$ in Eq.(\ref{eq:linear}) is set to project out from among $2^4$ states 
the $M=8$ states with eigenvalue $-1$ of $\hat h_l$. 
\par
We prepare a torus of $N_x\times N_y$ plaquettes ($N=2N_xN_y$), 
and wrap it by the spiral 1D MPS path running along the $x$-direction 
as shown in Fig.~\ref{faps3}(a). 
We add the MPS tensors one by one from top left to bottom right and determine their elements 
by performing a projection: 
the MPS tensor of the blue-colored site is determined by 
$Q_l$ giving $B_p=1$ on the upper plaquette to have PBC in the $x$-direction. 
The red-colored site is imposed $A_v~=1$ on the left and $B_p=1$ on the upper plaquette, 
the purple site has two $A_v=1$ and one $B_p=1$, 
and otherwise, no projection is given. 
To form a torus, the PBC in the $y$ direction is attained by diagonalizing the 
boundary operator, ${\cal H}^{\text{bd}}= -\sum_{j=1}^{N_x} (A_{v\in (j,N_y)} + B_{p\in (j,N_y)})$, 
consisting of operators on the last row. 
We find $D=4$ fold degenerate eigenstates with eigenvalue $-2N_x$. 
Finally, we apply two loop operators 
$Z_{{\rm h}}=\prod_{i \parallel x}\sigma^z_i, 
\;Z_{{\rm v}}=\prod_{i\parallel y} \sigma^z_i$, running along the 
horizontal and vertical closed paths, respectively, 
to classify the four degenerate ground states into topological sectors 
by their eigenvalues $\pm 1$. 
(For boundary and string operators see the third method in \S.\ref{sec:pbcmps}). 
\par
Figure~\ref{faps3}(b) shows the entanglement entropy(EE) $S_{N/2}$ for cutting the torus into half; 
given an extrapolation by an area of a cut (circumference) $N_x\rightarrow 0$, 
they show an exact extrapolation 
to the $-\ln 2$ value known as the topological EE of a ${\mathbb Z}_2$ spin liquid\cite{Jiang2012}. 
\ch{
Previous DMRG calculations on cylinders 
naturally select minimally entangled states, which are 
the superposed ones from the $\pm$ topological sectors\cite{Jiang2012,Jiang2013}. 
It is indeed generally difficult to obtain the topologically degenerate states of all sectors 
independently, as the eigenstates of the loop operators are difficult to identify\cite{Cincio2013,He2014}, 
and the entanglement entropy obtained by the variational ansatz have a tendency to be underestimated. 
Our method can elucidate them easily, and also {\it all the excited states exactly} 
by converting arbitrary sets of $\{A_v, B_p\}$ as $-1$. 
}
\par
Our MPS does not have TI as can be seen from the bond dimension and the EE when we divide the system 
into $n$ and $N-n$ sites in Figs.~\ref{faps3}(c) and \ref{faps3}(d) for 
four different topological sectors and \revt{ $N_x=2,3,4,5$, $N_y=12$, namely up to $N=120$.  
The maximum bond dimension required depends on $2N_x$, 
the maximum distance over which projection is not imposed, 
and we find $\chi_n\lesssim 400 N_x$ within the available range. 
The calculation is feasible up to $N_x=10-12$ for standard numerical resources, 
a size comparable to the 2D-DMRG. 
Notice that ED could only cope with spin-1/2 system up to $N\sim 30$. 
}
\par
\begin{figure*}
    \centering
    \includegraphics[width=18cm]{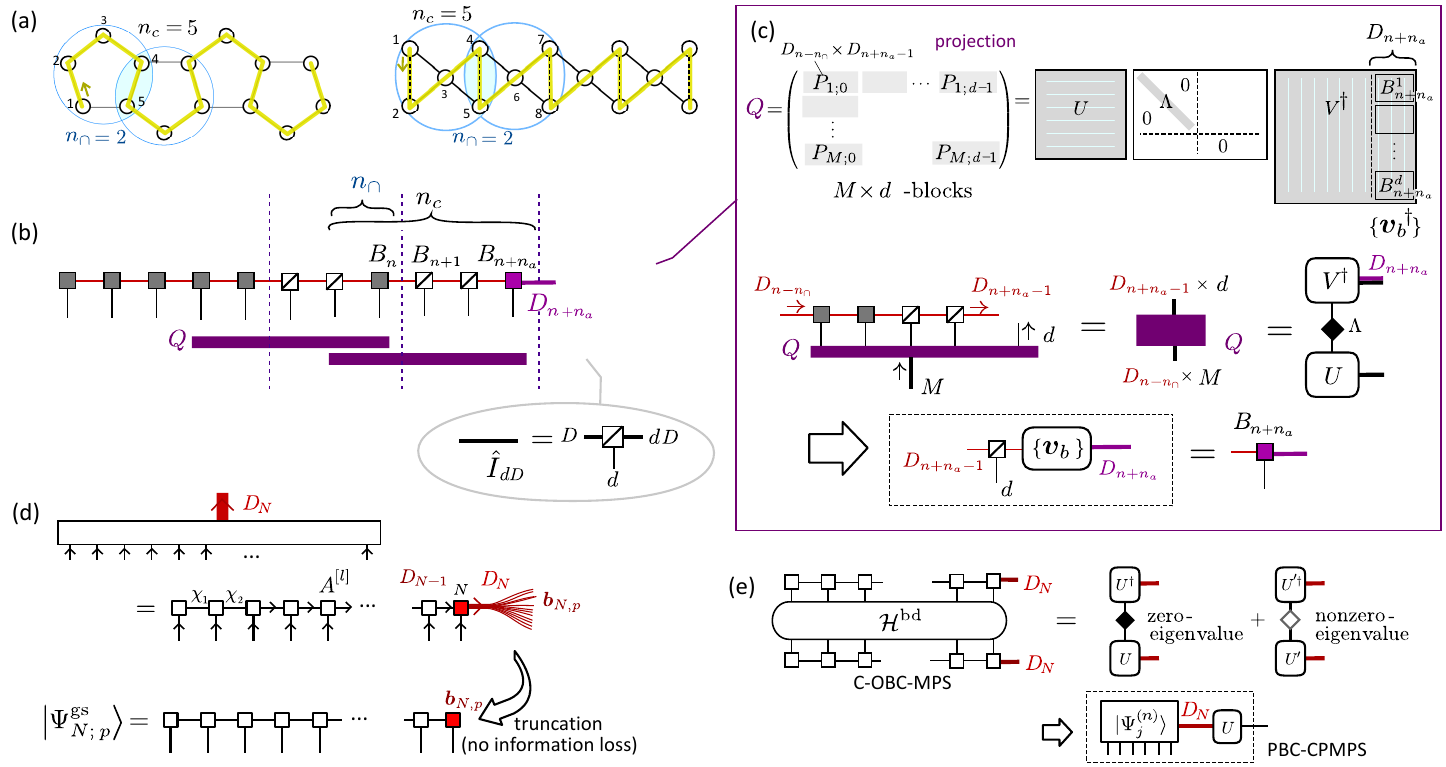}
    \caption{(a) Examples of lattices with $n_c=5, n_\cap=2$. 
    (b) Schematic illustration to construct the exact MPS solution with C-OBC. In and outward arrows indicate the order of determining the elements of $B_n$ 
    using the projection matrix $Q$. These elements are obtained by the singular value decomposition of $Q$. 
    (c) Process of determining $B_{n+n_a}$ at step 1 and 2 by the projection matrix $Q$, where we perform the SVD. 
    (d) Processes in step 4; 
       The set of $D_N$ states forms a left-normalized isometric tensor, 
       from which $p=1,\cdots, D_N$ independent $|\Psi_{N,p}^{\rm gs}\rangle$ are extracted. 
       The SVD is performed from right to left to reduce the bond dimensions to $\chi_n$. 
   \revt{ (e) Process of diagonalizing $\mathcal{H}^{\rm bd}$ to obtain a PBC-MPS from C-OBC-MPS. } 
     }
    \label{f5}
\end{figure*}
\section{Exact matrix product solutions}
\label{sec:exactmps}
In this section, we present the actual process of zeroing out part of the subspace of each cluster 
in determining the elements of the MPS tensor. 
This is done by successively adding the tensor and imposing projectors. 
The MPS is used not as a variational ansatz but as a convenient form of 
storing full information on the wave functions, 
whose advantage is to ``compress" the information without losing the exactness, 
relying on the canonical form and orthogonalization. 
\revt{
Before entering, we list the features of our exact MPS solution. 
\begin{itemize}
\item MPS tensors do not require TI form both for PBC and C-OBC, 
Whereas, the physical properties after averaging over degenerate states 
show TI for PBC, and the symmetry about the center for OBC. 

\item The exact ground states $|\Psi_N^{\rm gs}\rangle$ very often show degeneracies both for PBC and C-OBC, 
where the former solution is included in the latter. 
A full set of degenerate solutions is difficult to access by DMRG using MPS as a variational ansatz.

\item By restricting $N$ to the size comparable to DMRG, 
i.e. for which the bond dimension takes the realistic values, 
$\chi \lesssim O(10^4)$, the MPS ground states of the frustration-free Hamiltonians 
that has gapped ground states, gapless ground states, and long-ranged entangled states 
are obtained. 
\end{itemize} 
}

\subsection{Cluster-open boundary MPS in one dimension}
\label{sec:1d-cobcmps}
We now extend a protocol in \S.\ref{sec:clusterent} to construct the MPS wave function. 
The form of MPS for OBC is given as 
\begin{equation}
|\Psi_{\rm gs}\rangle= \sum_{\substack{i_1,i_2,\cdots,i_N\\ \alpha_1,\cdots,\alpha_{N-1}}} 
A^{[1]i_1}_{1\alpha_1} A^{[2]i_2}_{\alpha_1\alpha_2}\cdots A^{[N]i_N}_{\alpha_{N-1}1}\; 
|i_N\cdots i_1 \rangle ,
\end{equation}
where $A^{[n]i_n}_{\alpha_{n-1}\alpha_{n}}$ has a dimension $\chi_{n-1}\times \chi_n\times d$. 
We determine the size and elements of these matrices. 
\par
As a preparation, we choose a unit cluster consisting of $n_c$ sites 
and decide how to construct the lattice by 
making the neighboring clusters share $n_{\cap} (<n_c)$ sites. 
In Fig.~\ref{f5}(a) we show two example of the constructions of lattices with $n_c=5$ and $n_\cap=2$  
to guide the following explanation.  
We consider a 1D system with C-OBC, which is the OBC of clusters, not the sites. 
Compared to standard OBC, half of the interactions among $n_{\cap}$ sites belonging to 
$\hat h_1$ and $\hat h_{N_c}$ are lacking. 
We decide the 1D path of MPS on the lattice, e.g. as shown in bold lines in Fig.~\ref{f5}(a). 
\par
Next, we obtain a series of matrices $\{B_{\alpha_{n-1}\alpha_{n}}^{i_n}\}_{n=1}^N$ as shown in Fig.~\ref{f5}(b), 
which is written simply as $B_n$, 
and finally derive $A_n$ from $B_n$. 
These processes are performed following the steps given below. 
Further details are provided in the next subsection. 
\begin{itemize}
\vspace{-1mm}
\item[1.] Obtain $B_{1},\cdots, B_{n_c}$ as an initial set of MPS. 
We first diagonalize the $n_c$-site unit cluster, and from among the eigenstates, 
choose $M$ states $\{\xi^{m}\}$ to be projected out, 
expressed in the form of Eq.(\ref{eq:projctedst}), which give the element of $Q$. 
Here, $B_{1},\cdots, B_{n_c-1}$ is made using a set of unit matrices with bond dimensions $D_n=d^{n}$, 
so as to have nonzero coefficients for all $d^{n_c-1}$ basis states. 
Using $B_{1},\cdots, B_{n_c-1}$, we decide $B_{n_c}$ that 
projects out $\{\xi^{m}\}$ by the singular value decomposition (SVD) of matrix $Q$ (see Fig.~\ref{f5}(c)). 
The bond dimension of $B_{n_c}$ is $D_{n_c}=D_g=d^{n_c}-M$. 
\vspace{-2mm}
\item[2.] Start from $n$ sites and add $n_a\equiv n_c-n_\cap$ successive sites (one cluster)
to have $n'=n+n_a$ system. 
Again the $B_n$ of the first $n_a-1$ sites are made of unit matrices. 
When adding the last site, we again construct a matrix $Q$ that fulfills the linear equation 
for projection to the cluster ground state. 
The SVD of $Q$ will provide $B_{n'}$ of dimension $D_{n+n_a}$. 
\vspace{-2mm}
\item[3.] Repeat step 2 by setting $n'$ as updated $n$ until we reach the system size $n'=N$. 
The final matrix has dimensions $D_{N-1}\times D_N\times d$. 
A set of matrices $\{B_n\}_{n=1}^N$ represents $D_N$ degenerate solutions 
and form an isometric tensor. They are left-normalized (orthogonal) by construction. 
\vspace{-2mm}
\item[4.] We divide the matrix $B_N$ into $D_N$ columns. 
The $D_N$ independent solutions are these column vectors combined with 
$\{ B_n\}_{n=1}^{N-1}$ common to all of them. 
For each such set, we start from right $n=N-1$ toward $n=1$ and truncate the matrix. 
At each step, we divide the system into $n$ and right $N-n$ matrices 
and perform a Schmidt decomposition to discard the bonds that have zero Schmidt values 
which reduces the bond dimensions to $\chi_n$. 
$\{B_n\}_{n=1}^N$ is converted to $\{A_n\}_{n=1}^N$. 
\vspace{-2mm}
\end{itemize}
Notice that steps 4 is not necessarily needed 
because the tensors at step 3 
already serve as a full set of $D_N$ degenerate states. 
Step 4 offers the orthogonalized ground states separately 
and reduces their bond dimensions to the minimum; 
the truncation here does not lose information but shrinks 
the bond dimension by discarding the idle dimension. 
The redundancy in $D_n$ is because the matrices are shared with all degenerate states. 
In 1D system, we find that the truncation in step 4 
is needed for only the right half of the system, namely $\chi_n=D_n$ and $A_n=B_n$ for $n=1,\cdots N/2$. 
The states in steps 3 and 4 before and after the truncation 
are numerically precisely equivalent and both are exact. 
How the orthogonality of the degenerate MPS states 
is guaranteed is explained in detail in Appendix \ref{app:left-norm}. 
\subsection{\revt{Periodic boundary MPS}}
\label{sec:pbcmps}
The MPS that applies to the Hamiltonian with periodic boundary condition (PBC) takes the form 
\begin{equation}
|\Psi_{N}^{\rm pbc}\rangle=\sum_{\{i_n\}}^d  
{\rm tr} \big( \tilde A^{i_1}_1 \cdots \tilde A^{i_N}_N \big) |i_N\cdots i_1 \rangle. 
\label{eq:pbcmps1}
\end{equation}
There are several ways of constructing the PBC-MPS from C-OBC-MPS. 
\revt{Here, we explain the most practically useful one, which is diagonalizing the boundary operator. 
The other two methods are shown in Appendix \ref{app:pbc-mps}. }
\par
We first divide the Hamiltonian by the C-OBC term and the boundary term, given as
\begin{align}
\mathcal{H}_N = \mathcal{H}^{\text{C-OBC}}_N + \mathcal{H}^{\rm bd}.
\label{eq:hamedge}
\end{align}
Then we prepare a full set of OBC-MPS $\{|\Psi^{\rm gs}_{N,j}\rangle \}$ and diagonalize 
the boundary term, $\langle \Psi_{N,j}^{\rm gs}| \mathcal{H}^{\rm bd} |\Psi_{N,k}^{\rm gs}\rangle$, 
given in the form of $D_N\times D_N$ matrix as shown schematically \revt{in Fig.~\ref{f5}(e)}. 
The number of zero eigenvalues is the degeneracy of the PBC ground states, 
and we denote the $l$-th zero eigenvector as $(c_1^{(l)},\cdots, c_{D_N}^{(l)})$. 
The $l$-th PBC-MPS state is given as
\begin{equation}
|\Psi_{N}^{{\rm pbc}; (l)}\rangle = \sum_{j=1}^{D_N} c_j^{(l)} |\Psi_{N,j}^{\rm gs}\rangle,
\label{eq:pbcmps}
\end{equation} 
where using the column vectors of $B_N^{i}=(\bm b_{N,1}^{i},\cdots, \bm b_{N,D_N}^{i})$, 
the $N$-th tensor of the $l$-th PBC-MPS of dimension $\chi_{N-1}\times 1$ yields, 
\begin{equation}
\tilde B_N^{i}=\sum_{j=1}^{D_N} c_j^{(l)} \bm b_{N,j}^{i}. 
\label{eq:pbclastmat}
\end{equation} 
This method is the most efficient among the ones we developed (see Appendix \ref{app:pbc-mps}), allowing for larger $N$. 
The boundary operator is not restricted to boundaries but is used for the operators inside the lattice in 2D. 
\begin{figure}
    \centering
    \includegraphics[width=9cm]{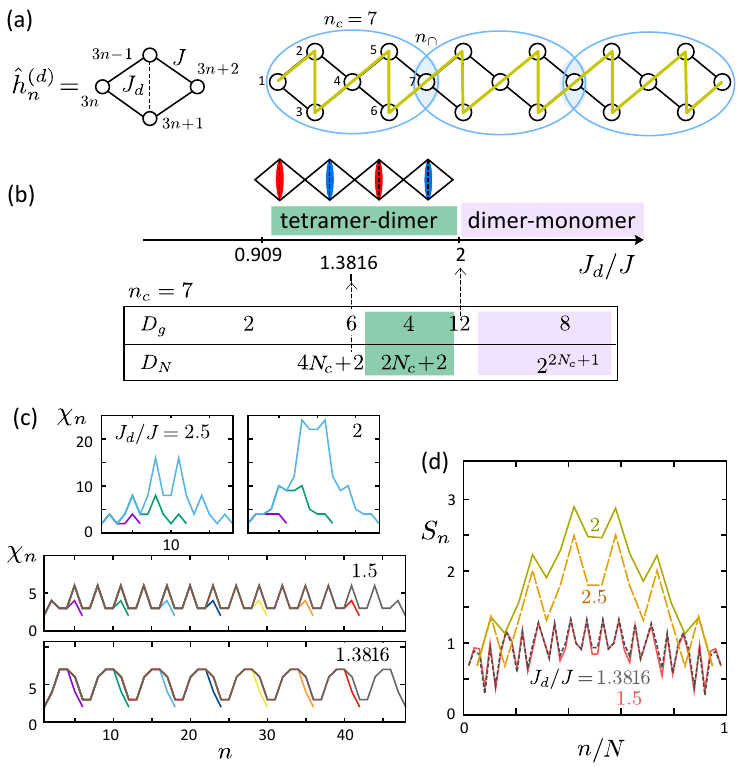}
    \caption{Application of MPS approach to the spin-1/2 Heisenberg diamond chain. 
    (a) Unit Hamiltonian and the lattice we adopt to construct MPS ($n_c=7$, $n_\cap=1$, C-OBC). 
    (b) The ground state phase diagram \cite{Takano1996} with two exact solutions, 
    tetramer-dimer and dimer-monomer states, with energies $E=n_v(e_t+e_s)$ and $n_ve_s$, 
    respectively, where $e_t=2(-J+J_d/8)$ and $e_s=-3J_d/4$ are the energies of 
    rung-triplet and rung-singlet state per diamond. 
    For $n_c=7$, $\hat h_l$ has several $D_g$ degeneracies. 
    The exact solutions are available at $J_d/J\ge 1.386$ with $D_g>2$.
    (c) Bond dimension $\chi_n$ for $J_d/J=1.3816, 1.5, 2, 2.5$ where 
    we take $N=7, 13, 19, \cdots, 45$. 
    (d) Bipartite entanglement entropy $S_n$ for corresponding data of panel (c), where we select $N=49$ and $19$, 
     which are the reasonable maximum size available for given $D_g$. 
     Data of (c) and (d) are averaged over all degenerate ground states. }
    \label{f6}
\end{figure}
\subsection{Example: spin-1/2 diamond chain}
To address the details of the C-OBC MPS protocol proposed in the previous subsection, 
we demonstrate the case of spin-1/2 diamond chain shown in Fig.~\ref{f6}(a). 
Two exact solutions for the Heisenberg model with two coupling constants, $J$ and $J_d$, are known\cite{Takano1994,Takano1996}, 
which are the tetramer-dimer and dimer-monomer states as shown in the phase diagram of Fig.~\ref{f6}(b). 
Deriving the corresponding MPS for these states refer to Case I in Fig.~\ref{f1}(a) 
where $\hat h_l$ is given {\it \'a priori}. 
The unit cluster to accommodate all these states without bias is $n_c=7$, namely two diamonds, 
which share $n_\cap=1$ with its neighbor. 
Notice that we may also choose $n_c=4$ (one diamond) and apply two different projections alternatively 
to simply obtain part of these states. 
\par
For each diamond we have a Heisenberg Hamiltonian $\hat h^{(d)}_{n}$ and the cluster Hamiltonian is given as
\begin{align}
\label{eq:hdiamond}
&\hat h_l= \hat h^{(d)}_{2l} + \hat h^{(d)}_{2l+1}, \nonumber\\
& \hat h^{(d)}_{n}=J (\bm s_{3n-1} + \bm s_{3n})(\bm s_{3n-2} + \bm s_{3n+1}) 
+ J_d\bm s_{3n-1}\bm s_{3n+1}.
\end{align}
We denote the down and up spin state $s_n^z=\mp 1/2$ for each site as $i_n=0$ and $1$, 
respectively, 
and describe the $n$-spin state as $|i_n i_{n-1},\cdots i_1\rangle$, e.g. 
$|0100\cdots\rangle$ given in the descending order of site indices. 
These states are indexed by $x=0,\cdots 2^n-1$ which is the base-ten numerals of these bits. 
For example, in diagonalizing Eq.(\ref{eq:hdiamond}) at $J_d/J=1.5$, 
we have $D_g=4$ fold degenerate lowest energy state, 
and we need to project out $M=2^{7}-4= 124$ states per cluster. 
\par
Let us explain the details of step 1 ($n=0$) and step 2 ($n>n_c$). 
At both steps, we add $n_a$-sites 
($n_a=n_c=7)$ for the initial step 1, and $n_a=n_c-n_\cap=6$ for successive step 2. 
Here, we need to obtain $B_{n+1},\cdots B_{n+n_a}$. 
For the first $n_a-1$ sites, the matrices describe the full set of basis $2^{n_a-1}$ equivalently, 
and are given by the disaggregation of a unit matrix, $\hat I$. 
By dividing the $2\times 2$ unit matrix into rows, 
we have $B_{n+1}^0=\hat I_{D_n} \otimes (1\:0)$ and $B_{n+1}^1=\hat I_{D_n} \otimes (0\:1)$. 
The second site matrices are given by dividing the $4\times 4$ as 
$B_2^0=I_{D_n} \otimes (1\: 0\: 0\: 0 ; 0\: 1\: 0\: 0)$ and $B_2^1=I_{D_n} \otimes (0\: 0\: 1\: 0 ; 0\: 0\: 0\: 1)$. 
We successively construct the rest of them up to $n+n_a-1$. 
\par
For $B_{n+n_a}$ we impose the condition to project out $\{|\xi^m\rangle\}$. 
The basis of $n_c$-site cluster is $|x\rangle =|i_{n_c} i_{n_c-1}\cdots i_1\rangle$, 
having $M$ different excited states, Eq.(\ref{eq:projctedst}). 
Next, for each of $m=1,\cdots,M$ different conditions, 
we prepare $D_{n-1}\times D_{n+n_a-1}$ matrix for $i_{n+n_a}=0,1$ 
\begin{equation}
P_{m ; i_{n+n_a}}\!=\!\sum_{x=0}^{2^{n_a-1}} 
\tilde \xi^{m }_{x} \big( B^{i_n}_{n} B^{i_{n+1}}_{n+1} \cdots B^{i_{n+n_a-1}}_{n+n_a-1} \big). 
\end{equation}
Using them we construct a projection matrix of $(M D_{n-1})\times (d D_{n+n_a-1})$, with $d=2$ as 
\begin{equation}
Q=\left(\begin{array}{ll}
P_{1;0} & P_{1;1} \\
\vdots \\
P_{M;0} & P_{M;1} 
\end{array}\right),\quad Q\: \bm v_b=0, 
\end{equation}
where $\bm v_b$ has dimension $(d D_{n+n_a-1})$. 
To obtain such $\bm v_b$ we perform a SVD as $Q=U\Lambda V^\dagger$, 
where $\Lambda$ has at most ${\rm min}(dD_{n+n_a-1},MD_{n-1})$ nonzero diagonal values. 
Using the $D_{n+n_a}$ rightmost columns of $V^\dagger$, 
which serve as $\bm v_b$, we operate them to the divided identity matrix and 
we obtain a set of matrices, $B_{n+n_a}^{0}$ to $B_{n+n_a}^{d-1}$, 
which are the $(dD_{n+n_a-1})\times D_{n+n_a}$ part of $V$ into $d$-blocks 
with $D_{n+n_a-1}\times D_{n+n_a}$. 
These processes are shown schematically in Fig.~\ref{f5}(c). 
\begin{figure*}
    \centering
    \includegraphics[width=17cm]{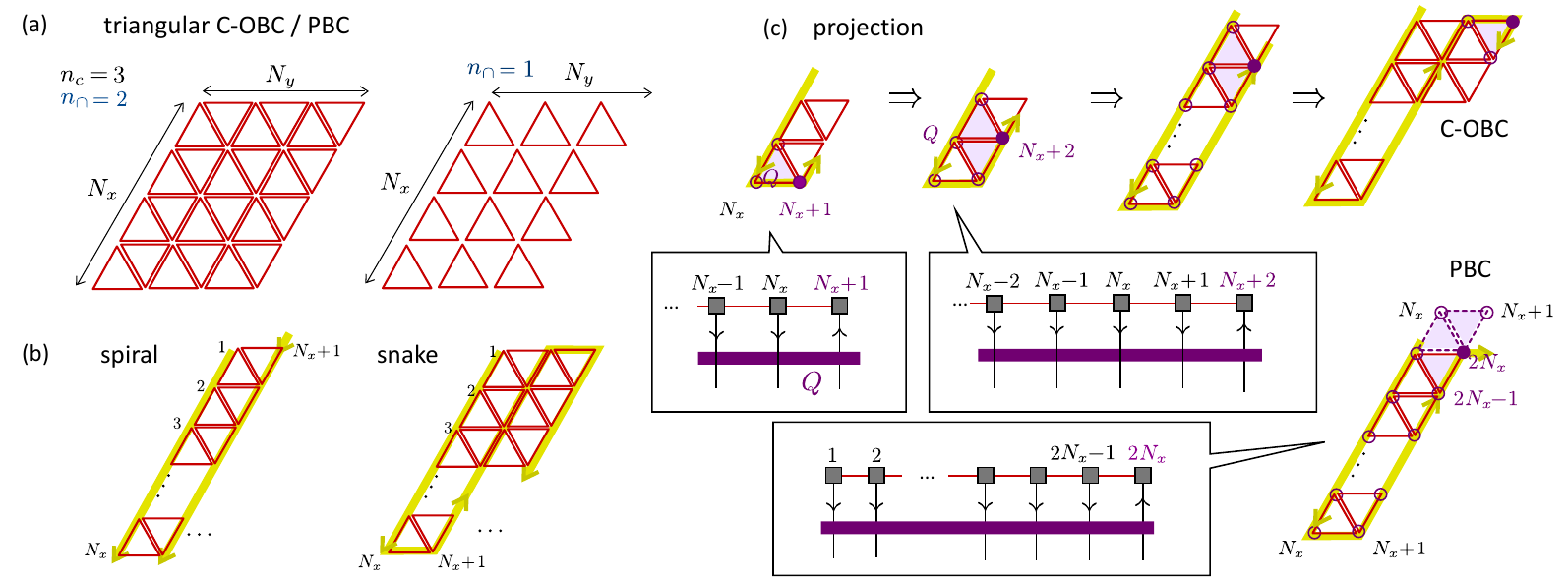}
    \caption{Construction of 2D MPS. 
    (a) Triangular lattice of $N=N_x\times N_y$ sites 
        using $n_c=3$ cluster with site overlaps, $n_\cap=2$ and $n_\cap=1$. 
    (b) Spiral and snake type 1D path for constructing MPS. 
    (c) How we operate the projection matrix $Q$ at each step for the snake type MPS and for PBC. 
        Bullet indicates the $n$-th site which we want to determine $B_n$ by the projection $Q$. 
        The matrix $Q$ includes several triangles highlighted in purple 
        where we want to exclude $\{\xi^m\}_{m=1}^M$, 
        and also includes the information of matrices of the open circles that are 
        not even included in the projected triangles. 
        When we apply PBC in the $x$-direction, we need to make projections 
        at $n=2N_x$ about three triangles with inputs from $2N_x-1$ sites 
        that share the first and second columns of the lattice. 
        In constructing a cylinder, we set the PBC in the $x$-direction that the snake runs. 
}
    \label{f8}
\end{figure*}
\par
We finally briefly explain the results obtained. 
Takano,{\it et al.} found the exact tetramer-dimer ground state at $J_d/J\lesssim 2$\cite{Takano1996}. 
On a single diamond, the triplet on a rung and the other triplet based on 
two sides entangle and form a tetramer, which has the energy $e_t=2(-J+J_d/8)$. 
On its neighbor, the singlet resides on a rung with energy $e_s=-3J_d/4$. 
The total energy of the product states of tetramer and dimer is given by 
$E=(e_t+e_s)n_v/2$ where $n_v$ is the number of rungs. 
Numerically, the ground state energy is found to extrapolate to $E/N$ smoothly with increasing $N$ 
at $0.909<J_d/J<2$. 
However, our method shows that there is another point, $J_d/J=1.3816$, 
not reported previously, that exhibits the energy crossing of the $n_c=7$ cluster Hamiltonian $\hat h_l$. 
One possibility is that $J_d/J=1.3816$ is a phase transition point. 
Below this point, $\hat h_l$ has $D_g=2$ and we cannot find an exact solution. 
Another possibility is that the trimer-dimer ground state 
is no longer an exact but an approximate ground state at $J_d/J<1.3816$. 
In such cases, a larger inter-cluster fluctuation including excited states has to be taken 
into account which is numerically hard to access. 
\par
In Figs.~\ref{f6}(c) and \ref{f6}(d), 
we show the bond dimension of MPS and the bipartite entanglement entropy 
$S_n$ when dividing the system into $n$ and $N-n$ parts. 
We find that the case of $J_d/J=1.3816$ and $1.5$ are similar in $S_n$ 
and are nearly flat, indicating the product-type ground state. 
However, the profile of $\chi_n$ differs much, which may suggest the change of 
the nature of the state below 1.3816. 
The other two cases with larger $D_g$ show a significant increase of $\chi_n$ and $S_n$. 
Our method can thus be a fingerprint of elucidating the nature of the phases. 
\par
The extension of the exact solutions of a diamond chain to 2D diamond lattice was reported\cite{Morita2016}, 
hosting a highly degenerate ground state with variants of the dimer covering pattern. 
It can also be dealt with in our framework by projecting it to the tetramer or dimer singlet states 
for $n_c=4$ diamond, and apply a 2D scheme in \S.\ref{sec:2d}. 
%
\begin{figure*}
    \centering
    \includegraphics[width=18cm]{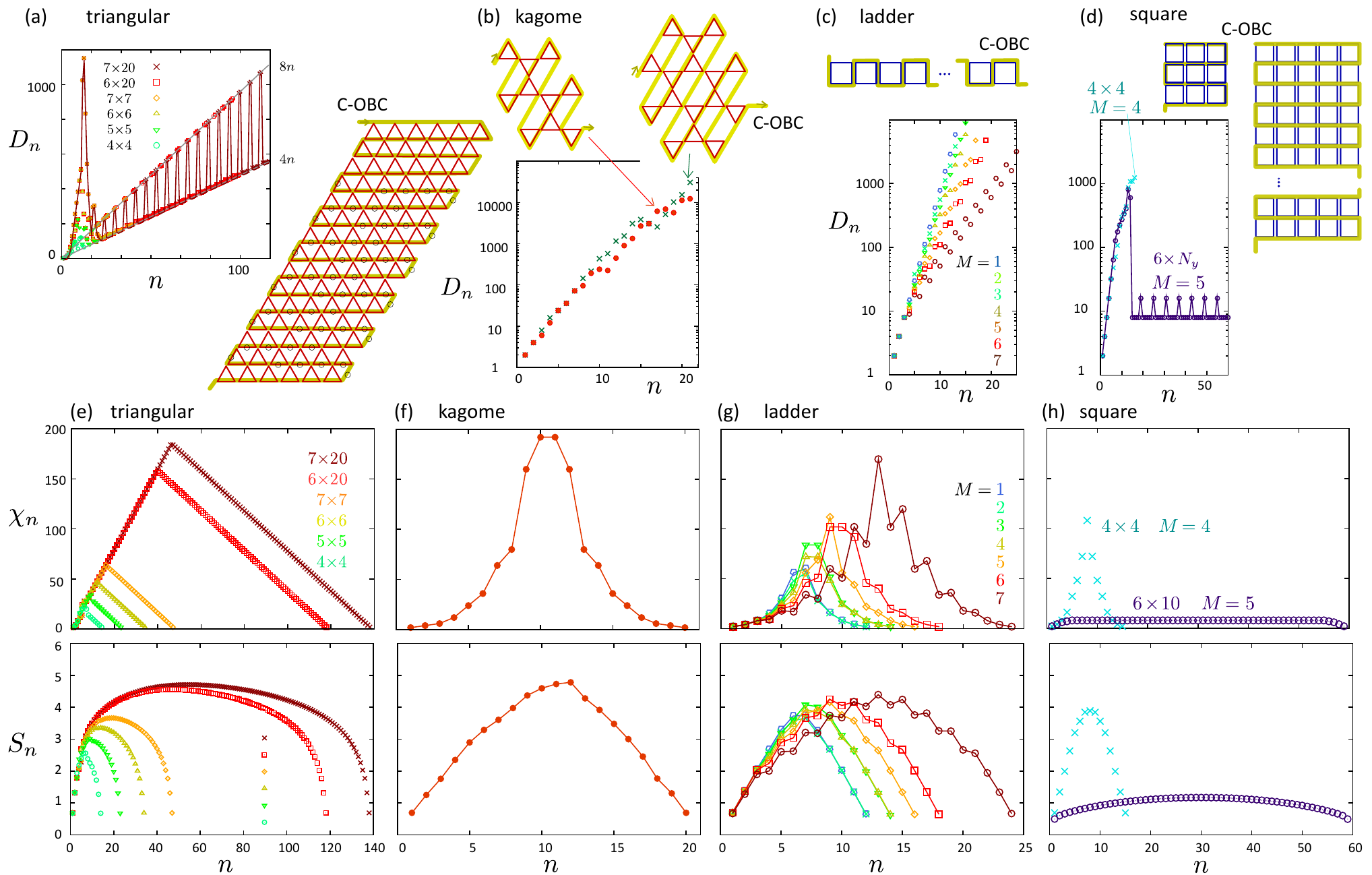}
    \caption{(a,b) $D_n$ of the exact MPS solutions of triangular and kagome lattice constructed
    using $n_c=3$ and $n_\cap=1$ for several choices of $N_x\times N_y$ with C-OBC. 
    The left panel in (a) is the case of $N_x=7$, where for $7\times 20$ system, we find $D_n=4n$ 
    for the bonds marked with a circle. 
    (c,d) Ladder and square lattice with $n_c=4$, $n_\cap=2$, C-OBC, where we change the number of sites $M$ 
    to be projected to see how $D_n$ grows. 
    (e-h) Bond dimension $\chi_n$ after truncating the MPS and EE, $S_n$ as functions of bipartition size $n$. 
    The data are averaged over $D_N$ ground states (see panels (a-d)), while for the kagome lattice, 
    we choose 500 ground states among $D_N=12288$ to save the computational cost
    which does not alter the quality of the result. 
   }
    \label{f10}
\end{figure*}
\section{Application to two-dimensional and random cases}
\label{sec:2d}
\subsection{Construction of two dimensional MPS}
The construction of MPS we showed in \S.\ref{sec:exactmps} is straightforwardly extended to 2D, 
while several points differ from 1D. 
For comprehensiveness, we consider the triangular unit $n_c=3$ as shown in Fig.~\ref{f8}(a): 
we can choose either $n_\cap=2$ and $1$ depending on how we design the boundary conditions and numerical costs. 
There are spiral or snake-type 1D paths of constructing MPS as shown in Fig.~\ref{f8}(b), 
while we here adopt the latter; 
At $n=N_x+1$ we only need to include a single triangle for projection. 
However, from $n=N_x+2$ to $2N_x-1$, we need to include two triangles highlighted and  
the input from the related sites marked with open circles. 
This is because the dimensions of matrices differ between sites that are separated 
by long distance over the MPS path, and we need to track the intervals to contract them. 
Besides the C-OBC, we can also construct a cylinder by taking PBC in the $x$-direction, 
in which case we need to project out the excited states of three triangles at $n=2N_x$ 
with input from $2N_x-1$ sites. 
The schematic illustration of constructing MPS is shown together in Fig.~\ref{f8}(c). 
\par
Using this setup, we construct the exact MPS with $n_c=3, n_\cap=2$ 
in Appendix \ref{app:tri-mps} and Fig.~\ref{f9}, finding a series of three-colored product state solutions. 
A similar state has been discussed as the exact ground states of the kagome lattice for 
the XXZ model at $J_z=-1/2$ \cite{Changlani2018} and the XYZ model. 
The present framework can detect the full set of degenerate solutions, 
and found the missing pieces of solutions different from the three-colored state, 
which were not characterized in these previous works.

%
\subsection{Variants of lattices for 2D MPS}
We now apply the method to several lattices to demonstrate the available size and constructions 
of the spin-1/2 lattice models, where we consider the case of $M=2$, $D_g=6$ unless otherwise noted. 
\par
{\it triangular lattice with C-OBC.}\; 
In Fig.~\ref{f10}(a) we show $n_c=3,n_\cap=1$ triangular lattice with C-OBC, 
where we plot $D_n$ as a function of $n$ for the MPS path taken along the yellow line in the snaky shape. 
For $7\times 20$ cluster with $N=139$ we find $D_N=560$ degenerate ground states. 
When $n<N_y$, we do not perform a projection so that the bond dimension grows in powers as $2^n$. 
At larger $n$ $D_n$ oscillates between four successive $4n$'s(see bonds are marked by a circle) and two $8n$'s, 
which are due to the periods of operating projectors. 
\par
{\it kagome lattice with C-OBC.}\; We apply the same treatment for the kagome lattice as in Fig.~\ref{f10}(b). 
Because the number of triangles that require a projection is much less 
than the triangular lattice, $D_n$ grows exponentially with $n$, which means 
that it is practically difficult to deal with this lattice. 
\par
{\it Ladder and square lattices with C-OBC.}\;
We now examine other types of unit clusters, $n_c=4$, $n_\cap=2$ in a spin-1/2 model. 
Here, we increase the number of sites to be projected out from $M=1$ to $7$ 
in the following order, 
\begin{align}
&|\xi_1\rangle =|0001\rangle+|0010\rangle+|0100\rangle+|1000\rangle, \nonumber \\
&|\xi_2\rangle =|1110\rangle+|1101\rangle+|1011\rangle+|0111\rangle, \nonumber \\
&|\xi_3\rangle =|0000\rangle,  \nonumber \\
&|\xi_4\rangle =|1111\rangle,  \nonumber \\
&|\xi_5\rangle =|0011\rangle+|0110\rangle+|1100\rangle+|1001\rangle, \nonumber \\
&|\xi_6\rangle =|0101\rangle +|1010\rangle,  \nonumber \\
&|\xi_7\rangle =|0001\rangle-|0010\rangle+|0100\rangle-|1000\rangle, \nonumber \\
&|\xi_8\rangle =|1110\rangle-|1101\rangle+|1011\rangle-|0111\rangle. 
\end{align}
Figure~\ref{f10}(c) shows $D_n$ for $M=1,\cdots,7$, finding that 
the exponential increase at $M\le 6$ is suppressed and when $M=8$ we are able to 
construct the exact ground state up to $10 \times 20$ lattice sites by suppressing 
the bond dimension to $D_n\le 800$. 
However, for the square lattice, the number of projections increases; 
In Fig.~\ref{f10}(d) we show the case of $M=4$ at $5\times 5$ and $M=5$ at $6\times N_y$, 
$N_y \sim 20$, where we find that $D_n \le 16$ and do not change much with $n$. 
At $M\ge 6$ the number of bases per square is too small to entangle the state, 
and we no longer find the exact solution. 
\par
{\it Entanglement entropy.}\hspace{2mm}
To make a more systematic understanding of the ground states of the above-mentioned lattices,  
we plot in Figs.~\ref{f10}(e)-(h) the bond dimension $\chi_n$ and the EE $S_n$, 
averaged over orthogonalized $D_N$ exact solutions with C-OBC. 
Here, when $\chi_n$ increases linearly as in the triangular case, 
the EE behaves as $\propto \ln n$, 
and its numerical cost is comparable to the well-known gapless 1D systems 
that the exact MPS feasible even in 2D, 
as one can see in the case of the ladder with quasi-linear $\chi_n$ in panel (g). 
The $M=5$ square lattice rather shows a 1D area-law-like behavior, that 
keeps $\chi_n$ constant over a wide range of the system. 
In contrast, the kagome lattice shows a volume law, 
$S_n \propto n$, that makes $\chi_n$ increase exponentially fast. 
\begin{figure}
    \centering
    \includegraphics[width=8cm]{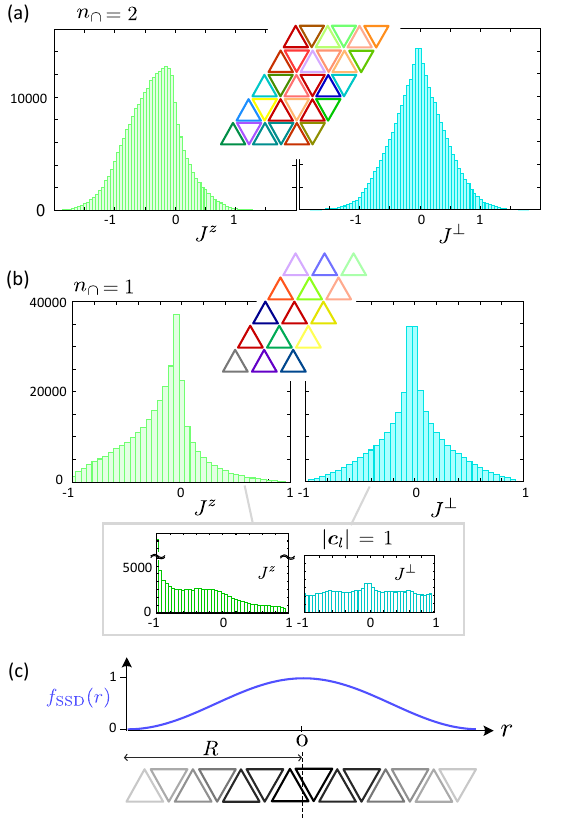}
    \caption{(a,b) Model with quenched randomness where $\hat h_l$ is determined by Eq.(\ref{eq:xirandom}) depends on $l$. 
    We set $n_\cap=2$ and $1$ where only the latter has a nontrivial ground state with $\sim 4N$ degeneracy. 
    the distribution of $J^z$ and $J^\perp$ over $N_x=N_y=300$ bonds are shown. 
    Lower panels are the distributions when we set $|\bm c_l|=1$ for all triangles. 
(c) Schematic illustration of the sine square deformation $f_{\rm SSD}(r)$ applied as envelope function 
    of the lattice Hamiltonian to vary the amplitudes of $\hat h_l$ gradually over the system. 
 }
    \label{f11}
\end{figure}
\subsection{Random systems}
We finally discuss the case of the spatially nonuniform Hamiltonians. 
\par
{\it Quenched bond randomness. }
Let us again consider the triangular unit cluster, $n_c=3$ of spin-1/2, 
where we may choose 
\begin{align}
&|\xi_{\Uparrow,l} \rangle = c_{l,1}|101\rangle + c_{l,2}|011\rangle + c_{l,3}|110\rangle ) \nonumber \\
&|\xi_{\Downarrow,l} \rangle =c_{l,1}|010\rangle+c_{l,2}|100\rangle + c_{l,3}|001\rangle), 
\label{eq:xirandom}
\end{align}
where $\bm c_l=(c_{l,1},c_{l,2},c_{l,3}) \in {\mathbb R}$ are random variables. 
The resultant $\hat h_l(\bm c_l)= |\xi_{\Uparrow,l} \rangle\langle \xi_{\Uparrow,l}|+ |\xi_{\Downarrow,l}\rangle\langle \xi_{\Downarrow,l}|$, 
is the XXZ cluster Hamiltonian whose three bonds typically have different values of $j^z$ and $j^\perp$.
Notice that if we take $\bm c_l\in {\mathbb C}$ while keeping the time-reversal symmetry, 
$\hat h_l(\bm c_l)$ starts to have the anisotropic and nonsymmetric 
exchange coupling terms such as $S_i^xS_j^y$ (see Appendix \ref{app:triangle}). 
\par
We now generate ${\bm c_l}$ from a uniform random distribution in the range $[-1:1]$ 
and consider two types of triangular lattice using the construction $n_\cap=2$ and $1$, 
as shown in Figs.~\ref{f11}(a) and \ref{f11}(b), respectively. 
The distribution of $J^z$ and $J^\perp$ for the two constructions are shown, both with a peak at zero. 
Unfortunately, when $n_\cap=2$ the system can host only trivial all-up and all-down state solutions 
with two-fold degeneracy at $N_x,N_y\ge 3$. 
When $n_\cap=1$, the degeneracy of the exact ground states increases to $\sim 4N$, 
which can be a good reference for the random-bond XXZ model, as it is the 
quantum version of the Edwards-Anderson model\cite{Edwards1975} studied extensively 
in the context of classical spin glass\cite{Binder1986}. 
Indeed, there is an increasing interest in trying to elucidate the quantum disordered phase 
\cite{Watanabe2014,Shimokawa2015,Liu2018,Han-Qing2019} in relevance to materials\cite{Kimchi2018}. 
\par
Quite remarkably, even if we vary the absolute values $|\bm c_l|>0$ arbitrarily 
while keeping its structure unchanged, 
the energy eigenstates of the triangles do not change so that the lattice ground state remains the same. 
However, the distribution of $J^z$ and $J^\perp$ change. 
We show in the lower panel of Fig.~\ref{f11}(b) the case where we normalize $|\bm c_l|=1$, 
which differs much from the unnormalized ones. 
The physical implication of the stability of the exact ground state is put forward for future studies. 
\par
{\it Sine square deformation. } We add one more example that the Hamiltonian is not spatially uniform but useful. 
We let the magnitude of $\hat h_l$ depend on the location of interactions as 
\begin{align}
&{\cal H}_{\rm SSD} = \sum_{l=1}^{N_c}  \hat f_{\rm SSD}(\bm r_{l}) h_{l}(\bm r_{l}), \nonumber \\
& f_{\rm SSD}(\bm r_{l})=\frac{1}{2}\big(1+\cos(\frac{\pi r_{l}}{R} )\big), 
\end{align}
where $R$ is the radius of a circle with its origin at the center of the lattice and 
$\bm r_{l}$ is the positional vector of the $l$-th cluster. 
We use the envelope function $f_{\rm SSD}$ which gradually scales down the interaction strength 
from 1 at the center to 0 at the edges (see Fig.~\ref{f11}(c)). 
This treatment is called sine-square deformation (SSD)\cite{Gendiar2009}. 
Solving ${\cal H}_{\rm SSD}$ gives the same ground state wave function as the case of PBC 
for the gapped system\cite{Hikihara2011,Katsura2011,Maruyama2011}. 
It is also found useful to reduce the finite size effect significantly\cite{Nishino2011,Kawano2022,Hotta2012,Hotta2013}. 
In the present framework, the deformation simply modifies the eigenvalues of $\hat h_l$ but since we 
set the lowest energy state to zero, the ground state solution does not change with this modification.

\section{Finding frustration-free models across the parameter space}
\label{sec:triangle} 
In this section, we demonstrate how to design a Hamiltonian (Case II) that can host 
the exact ground states by varying $\{\xi^m\}$ and $\hat h_l$ 
in \S.\ref{sec:tr1},\ref{sec:tr2} using the zigzag spin-1/2 chain as a platform. 
We find several unexplored frustration-free models and solutions. 
\subsection{Triangular unit}
\label{sec:tr1}
\begin{figure*}
    \centering
    \includegraphics[width=18cm]{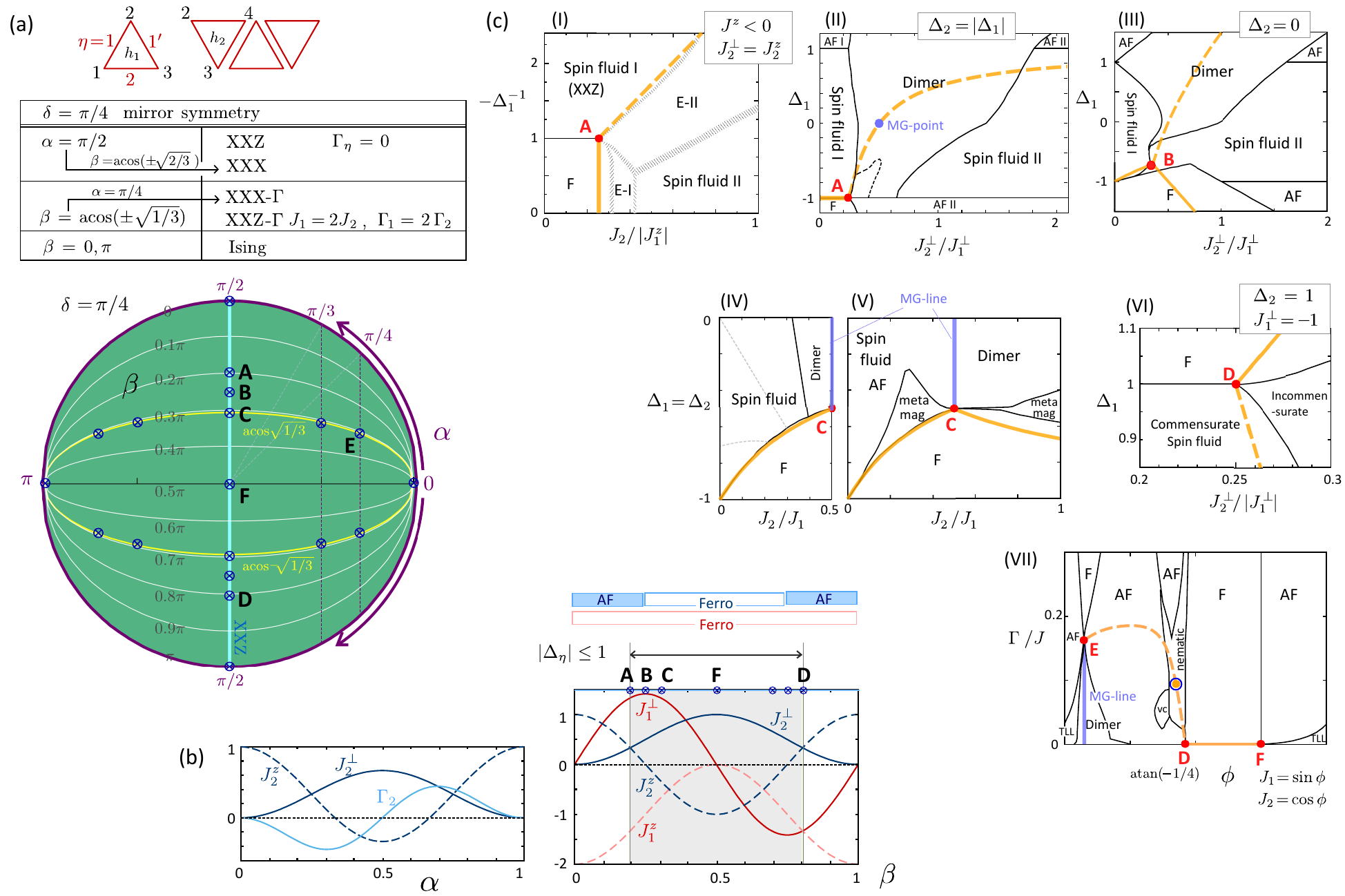}
    \caption{
(a) Diagram of solution space at $\delta=\pi/4$ (mirror symmetry) on the plane of $\alpha,\beta=[0:\pi]$. 
(b) Model parameters of Eq.(\ref{eq:zigzag}) along the XXZ line($\alpha=\pi/2$) 
$(J_1^\perp,J_1^z,J_2^\perp,J_2^z)=(\sqrt{2}\sin2\beta,-2\cos^2\beta,\sin^2\beta,\cos2\beta)$,
and $J_2=2J_1$ line ($\cos\beta=\sqrt{1/3})$ with 
$(J_2^\perp,J_2^z,\Gamma_2)=(2\sin^2\alpha/3,(2\cos2\alpha+1)/3, \sin 2\alpha/\sqrt{3})$. 
(c) Phase diagrams of previous studies (I)F-AF XXZ\cite{Plekhanov2010}, (II)F/AF-AF XXZ\cite{Somma2001}, 
(III)FXXZ-XY\cite{Jafari2007}, (IV,V)F-FXXZ \cite{Hirata2000,Gerhardt1998}, 
(VI)F-AF XXX\cite{Dmitriev2008}. 
(VII)XXX-$\Gamma$ models\cite{Saito2024-2}. 
Red bullets are the multicritical exact solutions with $M=2$ and 
Orange solid lines are the exact solutions with $M=4$ (broken lines are those that do not have 
solutions ($D_N=0$). 
The amplitudes of parameters are scaled together arbitrarily. 
%
%
}
    \label{f2}
\end{figure*}

Let us consider a unit triangle consisting of three spin-1/2 shown in Fig.~\ref{f2}(a), 
whose up and down spins are denoted as $i_n=0,1$ and its $d^{n_c}=2^3$ basis states 
are given as $|i_3 i_2 i_1\rangle=|000\rangle,|001\rangle,\cdots$. 
In a zero magnetic field, the system naturally requires a time-reversal symmetry, 
and from among four time-reversal pairs we choose a single pair for the penalty term as 
\begin{eqnarray}
|\xi_\Uparrow\rangle &=& \cos\alpha |000\rangle  + i \sin\alpha \big[\cos\beta|101\rangle 
\nonumber \\
&& \rule{12mm}{0mm}+\sin\beta \big(\cos\delta|110\rangle+\sin\delta|011\rangle \big) \big] 
\nonumber \\
|\xi_\Downarrow\rangle &=& \cos\alpha |111\rangle 
- i \sin\alpha \big[\cos\beta|010\rangle 
\nonumber \\
&& \rule{12mm}{0mm}+\sin\beta \big(\cos\delta|001\rangle+\sin\delta|100\rangle\big) \big], 
\label{eq:triangular_xi}
\end{eqnarray}
which are parameterized by $0\le \alpha,\beta,\delta \le \pi$, 
encompassing all possible realizations by the four basis states having the 
bond-symmetric interactions. 
To include the antisymmetric Dzyaloshinskii-Moriya terms \cite{Moriya1960}, 
we need to extend the parameter to complex variables which we present in Appendix \ref{app:triangle}. 
Unless otherwise noted, we focus on the case of $\delta=\pi/4$ 
that preserves the mirror symmetry of the triangle. 
\par
The local Hamiltonian is given as 
\begin{align}
\hat h_l&= |\xi_\Uparrow\rangle \langle \xi_\Uparrow| + |\xi_\Downarrow \rangle \langle \xi_\Downarrow| \nonumber \\
&=\!\sum_{\eta}\!\big( j_{\eta}^{\perp} (S^x_iS^x_j+S^y_iS^y_j)  + j_{\eta}^{z} S^z_iS^z_j 
+ \gamma_{\eta} (S^x_iS^y_j + S^y_iS^x_j) \big)
\label{eq:hl_tri}
\end{align}
where $\eta=1,1',2$ denote the three bonds forming a triangle, 
and we denote $j_\eta^{z}/j_\eta^{\perp}=\Delta_\eta$ by convention. 
The exchange interactions are
\begin{align}
j_1^{\perp}&= \sin^2\alpha\sin 2\beta \sin\delta, \nonumber \\
j_1^z \;&= \cos^2\alpha -\sin^2\alpha (\cos^2\beta-\sin^2\beta\cos2\delta),\nonumber\\
\gamma_1\:&=\sin2\alpha\sin\beta\cos\delta, \nonumber\\
j_{1'}^{\perp}&= \sin^2\alpha\sin 2\beta \sin\delta, \nonumber \\
j_{1'}^z \:&= \cos^2\alpha -\sin^2\alpha (\cos^2\beta+\sin^2\beta\cos2\delta),\nonumber\\
\gamma_{1'}&=\sin2\alpha\sin\beta\cos\delta, \nonumber\\
j_2^{\perp}&= \sin^2\alpha\sin^2\beta \sin 2\delta, \nonumber \\
j_2^z \;&= \cos^2\alpha +\sin^2\alpha \cos2\beta,\nonumber\\
\gamma_2\:&=\sin2\alpha\cos\beta.  
\label{eq:tri_jparam}
\end{align}
When $\tan\beta=\sqrt{2}$ (i.e., $\cos\beta=1/\sqrt{3}$), 
the triangle is $C_3$ symmetric and we find $j_1=j_{1'}=j_2$ and $\gamma_1=\gamma_{1'}=\gamma_2$. 
The value of $\alpha$ controls the anisotropy of spins 
and when $\alpha=\pi/2$, we find an XXZ type of interaction, 
$\gamma_\eta=0$. 
\subsection{Exactly solvable diagrams of zigzag chain}
\label{sec:tr2}
We consider a zigzag chain obtained by consecutively linking $(123), (234), \cdots$ clusters. 
\begin{align}
{\cal H}_N &= \sum_{\eta=1,1',2}\sum_{\langle i,j\rangle} 
\big( J_{\eta}^{\perp} (S^x_iS^x_j+S^y_iS^y_j)  + J_{\eta}^{z} S^z_iS^z_j 
\nonumber \\
& \rule{20mm}{0mm} + \Gamma_{\eta} (S^x_iS^y_j + S^y_iS^x_j) \big), 
\label{eq:zigzag}
\end{align}
where interaction runs over neighboring pairs of spins $\langle i,j\rangle$, 
and $J_1^{\perp/z}= 2j_1^{\perp/z}$, $J_2^{\perp/z}= j_2^{\perp/z}$, $\Gamma_1=2\gamma_1$, $\Gamma_2=\gamma_2$, 
because the diagonal bonds are doubled. 
\par
We first highlight the results obtained by this parameterization in Fig.~\ref{f2}(a). 
The ``solution-space diagram" is depicted by setting $\alpha$ as polar radius 
and varying $\beta$ in the vertical axis for fixed $\alpha$.  
All the parameters on this plane have exact solutions. 
There are two distinct lines; 
the vertical $\alpha=\pi/2$ line represents the XXZ model with $\Gamma_\eta=0$, 
and along the horizontal $\beta={\rm acos}(\pm\sqrt{1/3})$, 
all the diagonal rung bonds $\eta=1$ have twice as large amplitudes of interactions  
as those of the legs $\eta=2$ (see Fig.~\ref{f2}(a)). 
The representative points with high-symmetry model parameters are summarized in Table~II. 
\subsubsection{Spatially anisotropic XXZ models}
Let us first focus on the XXZ models ($\alpha=\pi/2$, see Fig.~\ref{f2}(a) and Table II(a)); 
The $\beta=0,\pi$ limits are the AF Ising chains coupled by the ferro(F) zigzag bonds, 
which has a trivial ground state. 
When $\cos\beta=\pm \sqrt{2/3}$ ({\bf A} and {\bf D}) the spin SU(2) symmetry is recovered 
and we find a ferro-antiferro(F-AF) zigzag Heisenberg (XXX) model. 
\par
The values of coupling constants as functions of $\beta$ are shown in 
the right panel of Fig.~\ref{f2}(b). 
The zigzag interaction is ferromagnetic ($J_1^z<0$), 
while both the F and AF leg interactions $J_2^z$ are realized. 
For $\Delta_\eta=J^z_\eta/J^\perp_\eta $ we find $|\Delta_\eta|\le 1$ between {\bf A} and {\bf D}. 
A wide range of magnetic anisotropy for a F/AF-F XXZ chain has room to afford exactly solvable ground states. 

\begin{table}[tbp] \label{tab1}
\caption{
Representative parameters $\alpha,\beta$ at $\delta=\pi/4$ plane with mirror symmetry. 
(a) $\cos\alpha=0$ ($\alpha=\pi/2$) line having $\Gamma_\eta=0$ and 
(b) $\cos\beta=\pm\!\sqrt{1/3}$ line with $J_1^{\perp/z}=\pm 2J_2^{\perp/z}$ and $\Gamma_2=2|\Gamma_1|$. 
These points have high symmetry in their model parameters, 
where XXX and I represent the Heisenberg and Ising model, XY the case with $J_\eta^z=0\:(\Delta_\eta=0)$. 
Points {\bf A}-{\bf F} refer to the ones that appear in Fig.~\ref{f2}.
}
\begin{tabular}{lrr rrrrcc}
(a) XXZ line & \multicolumn{3}{l}{$\cos\alpha=0$, $\delta=\pi/4$} \\
\hline
 &  $\cos\beta\;$ &  $(\:J_1^\perp,\;\;J_2^\perp,\;\;J_1^z,\;\;J_2^z\;\;)$ &  \\
\hline
\hline
F-AF I  &  $ \pm 1\:$  & $(\;\;0,\;\;\;\;0,\;\;\;\;-2,\;\;1\;\;)$ 
          & 
\rule{0mm}{3mm}\\
F-AF XXX  & $\sqrt{2/3}\:$  & $(\:4/3,\;1/3,-4/3,\;\:1/3\:)$ 
          &\rule{0mm}{3mm}&  {\bf A}  \\
F-AF XXX & $-\sqrt{2/3}\;$ &$\!(-4/3,\;1/3,-4/3,\;\:1/3\:)$
          && {\bf D} \\ 
\hline
F-F XXZ  & $\sqrt{1/3}\;$  & $(\:4/3,\;2/3,-2/3,-\!1/3\:)$ 
         && {\bf C} \\ 
FXXZ-XY  &  $\sqrt{1/2}\;$ & $(\:\sqrt{2},\;1/2,\!-1,\;\;\;\;0\;\;)$ 
         && {\bf B}  \\
FXXX(decoupled) & 0\;\; &   $(\;\;0,\;\;\;\;1,\;\;\;\;0,\;\;-1\;\;)$ 
         && {\bf F}  \\   
\hline 
(b) $J_1=2J_2$ line \rule{0mm}{6mm}&
 \multicolumn{3}{l}{$\cos\beta=\!\sqrt{1/3}$, $\delta=\pi/4$} \\
\hline
 &  $\cos\alpha$  & $(\;\;J_2^\perp,\;\;J_2^z,\;\;\;\Gamma_2\;\;)$ &   \\
\hline
F-F XXZ & 0 \;  & $(\;2/3,\;\;-\!1/3,\;\;\;0\;\;)$ & \\
XY$\Gamma$ & $\pm 1/2$  & $(\;1/2,\;\;\;\;\;0,\;\;\pm 1/2\;)$  &  \\
AF XXX-$\Gamma$  & $\pm\!\sqrt{1/2}$ & $(\;1/3,\:\;1/3,\pm \sqrt{1/3}\;)$ 
        && {\bf E} \\ 
AF I & $\pm 1\;$   & $\;(\;\;0,\;\;\;\;\;1,\;\;\;\;0\;\;)$  & \\
\hline
\end{tabular}
\end{table}
\par
The present framework successively connects different exact solutions that appear in different models 
that are separately studied in previous literature. 
The six phase diagrams that host {\bf A} to {\bf D} are summarized in 
Fig.~\ref{f2}(c) (I)-(VI)\cite{Plekhanov2010,Somma2001,Jafari2007,Hirata2000,Gerhardt1998,Dmitriev2008,Saito2024-2}, 
which are variants of F-AF XXZ models. 
Here, we refer to F or AF interactions as those of $J_\eta^z$, 
because the local unitary transformation rotating the spin quantization axis on either of the legs by $\pi$ 
converts $J_1^\perp\rightarrow -J_1^\perp$, but does not change the physical states. 
It is widely accepted that a series of materials consisting of edge-shared CuO$_2$ chains 
such as Rb$_2$Cu$_2$Mo$_3$O$_{12}$\cite{Hase2004}, Na$_2$Cu$_2$O$_2$\cite{Drechsler2006}, 
and LiCuVO$_4$\cite{Enderle2005} are represented by the F-AF XXZ models. 
\par
In the phase diagram Fig.~\ref{f2}(c)-(I) taken from the DMRG study\cite{Plekhanov2010}, 
two gapless spin fluid phases are observed, having a zigzag and strong-leg characters, respectively. 
The ferromagnetic (F),  spin fluid I, and the other two phases with massive excitations 
merge at the single point {\bf A}, which forms a multicritical point. 
This point is exactly solved in our framework. 
\par
Analogous XXZ-types of models that interpolate 
F-AF XXX $(\Delta_1=-\Delta_2=-1)$ or FXXZ-XY $(\Delta_1=-1/\sqrt{2}, \Delta_2=0)$
zigzag interactions in Figs.~\ref{f2}(c)-(II)\cite{Somma2001} and (III)\cite{Jafari2007}
exhibit essentially similar diagrams, insensitive to $\Delta_2$. 
At $J_2^\perp/J_1^\perp \gtrsim 0.5$, 
one finds a gapped dimer singlet phase, sandwiched by Spin fluid I and II. 
When $|\Delta_\eta|\ge 1$ the system becomes massive and develops either F or AF long-range orders. 
\par
It is noteworthy that in diagram (III), Ref.[\onlinecite{Jafari2007}]
did not detect the multicritcal point {\bf B} located at $(J_2^\perp/J_1^\perp,\Delta_1)=(\sqrt{1/8},-\sqrt{1/2})$. 
This manifests that for model parameters with an irregular anisotropy, 
it is difficult to predict or confirm numerically whether several phases 
exactly meet at a single point. 
\par
Similarly, in the AF-AF SU(2) Heisenberg case (diagrams (IV, V)), 
there is another point {\bf C} with $J_2/J_1=0.5$, $\Delta_1=\Delta_2=-1/2$ 
at which the F and Dimer phases meet. 
As we see shortly, it is the endpoint of the Majumdar-Ghosh line, 
hosting an exact dimer singlet-product ground state\cite{Majumdar1969,Majumdar1969-2}. 
The phase diagram in panel (V)\cite{Gerhardt1998} was studied in search of the metamagnetic phase, 
which shows a magnetization jump in an applied magnetic field as observed in 
several materials like Fe$_x$Mn$_{1-x}$TiO$_3$, GdNi$_2$Sb$_2$ and GdCu$_2$Sb$_2$ 
\cite{Ito1992,Lelievre-Berna1993,Kaczmarska1995}. 
Indeed, {\bf C} is a multicritical point where the metamagnet merges with dimer and F phases. 
\par
The multicritical point {\bf D} in the phase diagram (VI) has further interesting properties; 
it has both the $S=0$ ground state and fully polarized $S=N/2$ ferromagnet to be perfectly degenerate, 
and the former is not the Majumdar-Ghosh product state but 
another highly entangled state called uniformly distributed resonating valence bond state (UDRVB). 
Hamada, {\it et.al.} discovered a UDRVB as an exact ground state 
at $J_2=-J_1/4$, $(J_1<0)$ (this F-AF model is called the generalized railroad trestle model)\cite{Hamada1988}. 

\subsubsection{$J_1=2J_2$ $\Gamma$ model}
At $\alpha\ne \pi/2$, the finite $\Gamma_\eta$-terms appear. 
We particularly focus on the $\cos\beta=\sqrt{1/3}$ line where the $C_3$ symmetry of the 
unit triangle leads to twice as large zigzag coupling against the leg couplings, 
$J^{\perp/z}_1=2J_2^{\perp/z}$ and $\Gamma_1=2\Gamma_2$. 
When we further take $\cos\alpha=\sqrt{1/2}$ the AF Heisenberg coupling, 
$J^{\perp}_\eta=J^{z}_\eta\equiv J_\eta$ is realized at {\bf E}. 
It appears as a multicritical point of the phase diagram of the Heisenberg $J_1$-$J_2$ model 
with finite $\Gamma$ term shown in Fig.~\ref{f2}(c)-(VII)\cite{Saito2024}. 
Here, we plot it on the plane of $\Gamma_\eta/J_\eta$ and $\phi={\rm atan}(J_2/J_1)$ 
with $J_2=\cos\phi$ and $J_1=\sin\phi$. 
The multicritical point {\bf E} has $J_1=2J_2$ and $\Gamma_\eta=\sqrt{3}J_\eta$, 
where five phases meet, and similarly to point {\bf C} it is part of 
the Majumdar-Ghosh line that extends from $\Gamma_\eta=0$. 
Our method interpolates point {\bf E} with the other two points, 
{\bf D} and {\bf F} within the same $\delta=\pi/4$ (Fig.~\ref{f2}(a)). 
Point {\bf D} hosting RVB appears at $\phi={\rm atan}(-1/4)$ where the RVB with $S=0$ and the 
fully polarized ferromagnetic solution $S=N/2$ coexist. 
The latter extends to larger $\phi$ toward the endpoint {\bf F}, at which we find a decoupled ferromagnetic Heisenberg chain. 
\begin{figure}
    \centering
    \includegraphics[width=7cm]{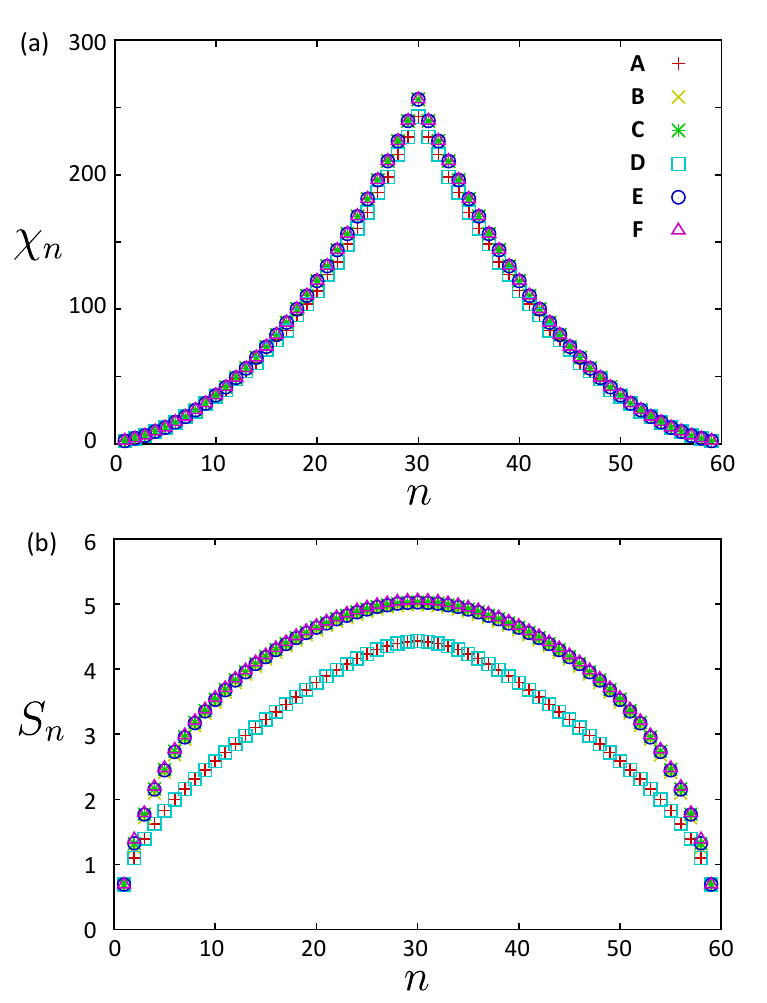}
    \caption{(a) Bond dimension $\chi_n$ and (b) EE $S_n$ of the exact MPS solutions of the zigzag spin-1/2 chain 
  with $N=60$ C-OBC, obtained at parameters {\bf A}-{\bf F} in Fig.~\ref{f2}(a). 
  The results are averaged over the orthogonalized $D_N$-degenerate solutions. 
}
    \label{fn12}
\end{figure}
\subsubsection{Exact MPS solutions and degeneracy}
Here, we obtain an exact MPS solutions for the the parameters {\bf A}-{\bf F}, 
using the protocol explained in \S.\ref{sec:exactmps}. 
The degeneracies follow
\begin{align}
    D_{N}^{\text{C-OBC}}=\left\{ 
    \begin{array}{ll}
    (N+2)^2/4 & ({\rm even}\;\;N)  \\ 
    (N+1)(N+3)/4 \;\;&  ({\rm odd}\;\;N). 
    \end{array}
    \right.
    \label{eq:DN_TOBC_square}
\end{align}
Figure~\ref{fn12} shows the bond dimension $\chi_n$ and EE $S_n$ for $N=60$ C-OBC solutions. 
One finds that $\chi_n$ after truncated is basically equivalent to $D_n$ up to 
$n\le N/2$ and the EE follows $S_n\propto \ln n$. 
\par
Off these points, there are other types of degeneracies within the parameter space of Eq.(\ref{eq:triangular_xi}), 
which are summarized in Appendix \ref{sec:tr4}. 
One interesting aspect is that there are many different models that have 
solutions following the same $D_N$. 
It is natural to expect that there are underlying physics in common, which should 
be one of the future perspectives. 
\\
\subsection{Exact solution lines in the phase diagrams}
\label{sec:tr3}
So far we have designed the frustration-free models following Case II in \S.\ref{sec:clusterent-1}. 
These solutions appears as ``isolated points", {\bf A}-{\bf F}, 
in the phase diagrams of Fig.~\ref{f2}(c). 
However, if we examine a given Hamiltonian in each diagram and vary the model parameters, 
we find that these points are not isolated but are connected to the exact solution lines. 
\par
To find such lines, we diagonalize the unit cluster Hamiltonian $\hat h_l$, 
and see how the degeneracy $D_g$ of the lowest energy manifold changes with model parameters.  
In \S.\ref{sec:tr1} we considered $D_g=6$ ($M=2$) 
for multi degenerate solutions at {\bf A}-{\bf F}, which are the tricritical(multicritical) points. 
In addition, $\hat h_l$ can have $D_g=4$ ($M=4$) along the bold lines of Fig.~\ref{f2}(c)(I)-(VII). 
Among them, the solid lines have bulk exact solution, 
which include the Majumdar-Ghosh lines or points, 
which is the singlet-product ground state of the zigzag chain. 
It is noteworthy that these exactly solved lines 
very often coincide with the numerically obtained phase boundary, 
e.g. in Fig.~\ref{f2}(c)-(IV),(V), at which F and spin fluid phases meet. 
The broken lines do not have such solutions but connect different multicritical points. 
Indeed, in Fig.~\ref{f2}(c)-(VII), we newly find an exact nematic product solution 
on the broken line inside the nematic phase. 
This broken line connects the two multicritical points, {\bf E} and {\bf D}. 
These results show that the present framework is useful to clarify the ground state phase diagram 
with a very small numerical cost of diagonalizing the small unit cluster. 
It reminds us of a level spectroscopy analysis that successfully detects 
the phase boundary using exact diagonalization\cite{Nomura1994,Kitazawa1997}. 
\par 
These exact solutions discussed here in Fig.~\ref{f2} belong to the gapless ground states 
of the frustration-free Hamiltonian. 
Indeed, the ones found along the XXZ lines are equivalent 
to the one that is proven to be gapless based on the exact analysis using anyons\cite{Batista2009}. 
\par
\begin{figure}
    \centering
    \includegraphics[width=8.5cm]{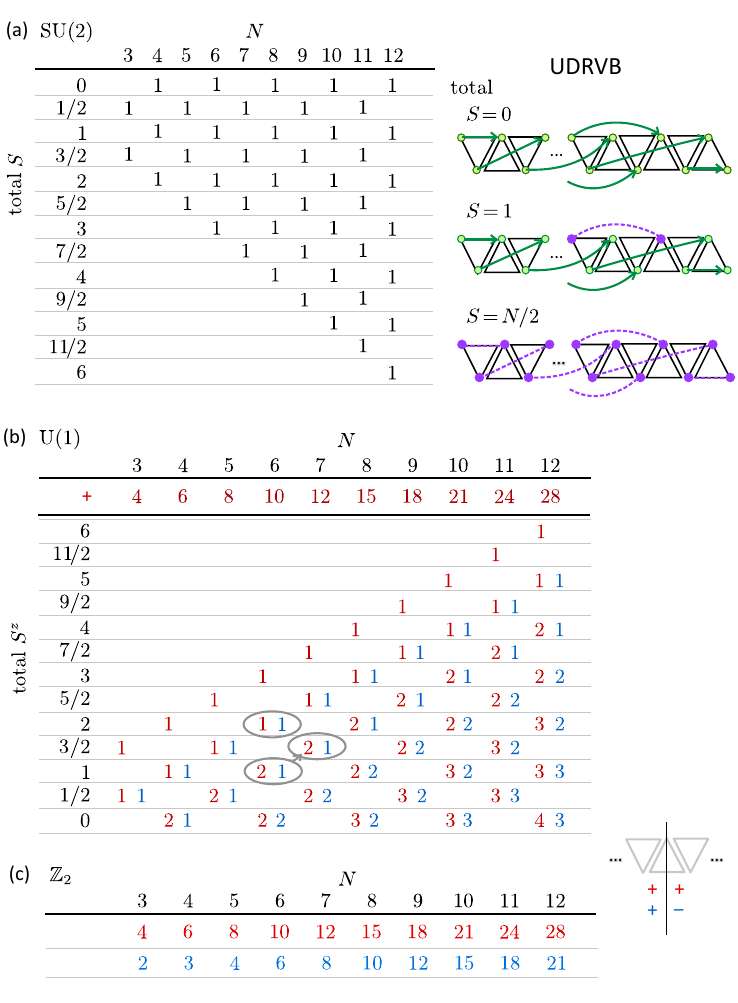}
    \caption{Distribution of degeneracy of $N^2$-type ground states of the zigzag chain with C-OBC 
classified according to the conservation number. 
(a) Heisenberg model with SU(2) symmetry realized at $\beta={\rm acos}(-\sqrt{2/3})$ at point {\bf D}. 
A schematic illustration of the RVB state is shown on the right panel where 
a pair of sites connected by arrows denotes the singlet, and a pair connected by a broken line is the triplet. 
(b) XXZ model ($\alpha=\pi/2,\;\delta=\pi/4$) with U(1) symmetry, 
classified by the total $S^z$ where we show only the $S^z \ge 0$ part. 
(c) XYZ model at $\alpha\ne \pi/2$ having ${\mathbb Z}_2$ symmetry. 
The red and blue numbers in panels (b) and (c) are those that have $+1$ and $-1$ eigenvalues about 
the mirror operation against the vertical axis of the center of the cluster. 
}
    \label{f4}
\end{figure}
%
\subsection{Symmetry and RVB state}
\label{sec:udrvb}
The exact solutions for a given $N$ can be classified by the symmetries. 
Here, we focus on the zigzag chain with C-OBC (see Fig.~\ref{f2}(a) with $\delta=\pi/4$), 
which has $D_N$ given in Eq.(\ref{eq:DN_TOBC_square}). 
The spins have SU(2) symmetry at {\bf A}, {\bf D}, {\bf F}, 
U(1) symmetry at other points along the XXZ line, 
and ${\mathbb Z}_2$ the other parts of the $\alpha$-$\beta$ plane 
except for a few points. 
\par
To understand the origin of degeneracy, 
it is useful to classify the solutions according to the conservation numbers. 
This is done by projecting the solutions to the corresponding subspace. 
In Figs.~\ref{f4}(a)-(c) we show the numbers of solutions 
for a given $N$ on each subspace, 
where total $S$ and total $S^z$ are used for SU(2) and U(1) cases, respectively. 
We classify the U(1) and ${\mathbb Z}_2$ cases further by their eigenvalues $\pm 1$ 
about the mirror operation against the plane perpendicular to the zigzag ladder. 
\par
We now show that the present analysis helps us clarify the nature of the RVB state at point {\bf D}. 
The exact solution is called RVB given in the form\cite{Hamada1988}, 
\begin{align}
&|\Phi_{\rm RVB}\rangle= \sum_{i_1<j_1,i_2<j_2,\cdots}[i_1,j_1][i_2,j_2]\cdots, \nonumber\\
& \quad [i,j]=(|0 1 \rangle -|1 0\rangle) /\sqrt{2}, 
\label{eq:udrvb}
\end{align}
where the summation is taken over all possible combinations of pairs of sites with no duplication on site indices. 
Here, $[i,j]$ is the singlet formed on $i,j$-th pair of sites, 
written as arrows in Fig.~\ref{f4}. 
It was shown that\cite{Hamada1988} the $S=0$ RVB is degenerate with the $S=N/2$ fully polarized state 
for PBC because this parameter is at the boundary of the three phases (see Fig.~\ref{f2}(c)-(VI)). 
A new finding here is that in C-OBC, 
all the spin sectors (that were the excited states in PBC) become degenerate with $S=0,N/2$ and join the ground state; 
we find that the number of degeneracy in Eq.(\ref{eq:DN_TOBC_square}) comes from 
the number of all different total-$S^z$ and total-$S$ sectors given in Fig.~\ref{f4}(b).  
Using our MPS solution, we discover the exact form of the $S=1$ state: 
it is obtained by replacing the first singlet $[i,j]$ to triplet $(i,j)$ as
\begin{equation}
 |\Phi_{S=1}\rangle=  \sum_{i_1<j_1,i_2<j_2,\cdots}  (i_1,j_1)[i_2,j_2],[i_3,j_3]\cdots. 
\label{eq:udrvb-magnon}
\end{equation}
In the same manner, the $S=m$ state has the form of replacing the first $m$ successive singlet 
in Eq.(\ref{eq:udrvb}) to triplet as well. 
The reason why the states $S\ne 0, N/2$ become the excited state for PBC is that 
they cannot fulfill the translational invariance. 
It is natural to expect all of them to collapse to the ground state at $N\rightarrow \infty$, 
providing us with the state beyond the simple RVB, which needs further clarification.

\section{summary and discussion}
We have proposed a framework to concurrently design a frustration-free model 
and obtain its exact ground state on a sufficiently large finite-size cluster in quantum many-body systems. 
\revs{Even knowing whether the candidate model is frustration-free or not is generally difficult as 
it is the quantum $k$-QAT problem known as the QMA$_1$-complete in the numerical complexity theory. 
Our framework addresses a pragmatic challenge to this issue and succeeded in 
not simply verifying this question but in providing an actual form of their exact solutions, if it is present.}
\par
We begin by introducing a unit cluster and categorize its local Hilbert space into two distinct manifolds: 
one contributing to the ground state and the other not. 
The local cluster Hamiltonian is designed to have the former the lowest(zero) energy and the latter 
the excited energy, and 
the lattice Hamiltonian is formulated as their summation. 
By zeroing out the excited manifold, 
the lowest-energy manifold enters the full many-body lattice wave function. 
\revt{If such wave function exists, the lattice Hamiltonian is frustration-free.}
\par
It is then imperative to entangle the lowest energy states among different clusters. 
The key to our protocol projects the Hilbert space onto these states for all clusters, 
ensuring an unbiased ground state determination that can be systematically applied across various models. 
When further employing the MPS approach, 
we can progressively expand the system size and iteratively apply the projection 
to newly added clusters to determine their matrices. 
We can take full advantage of the techniques developed for MPS like canonical form or truncation, 
while abandon the typical algorithm for variational ansatz. 
Given the precise knowledge that the ground state energy is set to zero and the positive semidefiniteness 
of the Hamiltonian, validating exact solutions is straightforward. 
\par
\revt{
One can systematically search for such Hamiltonians by parameterizing 
the choice of lowest energy cluster states. 
This approach offers various types of models and facilitates connections 
between different exact solutions found in various contexts. }
We provide a demonstration using the spin-1/2 zigzag chain, 
showcasing exact ground states and 
revealing connections between previously reported phase diagrams in anisotropic XXZ-type models. 
As a novel discovery, we found that the resonating valence bond (RVB) state of 
the ferro-antiferro zigzag chain model 
previously reported by Hamada {\it et. al.} \cite{Hamada1988} 
exhibits additional macroscopic degeneracy across all total $S$-sectors. 
This implies the condensation of all allowed numbers of magnons, 
suggesting a multi-magnon Bose-Einstein condensate atop the RVB sea. 
\par
Related work by Batista {\it et al.} \cite{Batista2009,Batista2012} 
has found an exact solution for the antiferromagnetic XXZ model on the zigzag ladder, 
exhibiting the same degeneracy as shown in Eq.(\ref{eq:DN_TOBC_square}). 
They employed a generalized Jordan-Wigner transformation to map spins to anyons 
and identified a BEC ground state of anyons carrying momentum $Q$, represented as 
$(a_Q)^n(a_{-Q})^m|0\rangle$, 
where the choice of adding $n$ and $m$ anyons explains the origin of the degeneracy. 
It corresponds to the XXZ line of our zigzag chain at $\alpha=\pi/4$ with total $S^z=n+m-N/2$. 
They also derived the specific form of the MPS solution for $Q=2\pi/p$ of bond dimension $p-1$. 
However, there is no one-to-one correspondence between their periodic MPS representation and ours, 
as the MPS representation has large facultativity. 
Our method helps to relate these results with other models as we did in Fig.~\ref{f2}. 
\par
We also demonstrated the application to 2D systems within a comparable computational cost to the 1D case. 
However, in 2D, the bond dimensions can grow more rapidly than 1D, 
especially when the overlap of neighboring clusters $n_\cap$ is small. 
Notably, on the kagome lattice, the degeneracy undergoes exponential growth, 
while nonetheless, intriguing exact MPS solutions exist; 
in a spin-1/2 kagome lattice model at specific interaction strength, 
$J^z/J^\perp=-1/2$, called XXZ0, a manifold of three-coloring product states\cite{Changlani2018} 
is detected via exact diagonalization (ED). 
The XYZ model also showcases three-coloring degenerate ground states\cite{Palle2021}. 
They feature macroscopic degeneracy attributed to classical frustration effects giving three-coloring patterns
\cite{Harris1992,Chalker1992,Henley2009}, 
whereas there exist more ground states that defy such simple explanations. 
Our framework can capture even such missing states, and can further 
test many other possibilities proposed as presumable candidates. 
\par
\revt{
We briefly refer to a series of works on injective MPS having a translationally invariant(TI) form\cite{Perezgarcia2007,Cirac2021}. 
It is mathematically proven that for any given injective MPS one can find a frustration-free parent Hamiltonian 
and that its ground state is unique and gapless. 
Contrastingly, our MPS is the first to abandon the TI which is obtained by the systematic and general protocol. 
By taking pragmatic advantage of non-TI, one can restrict the system size up to which 
the numerical resource allows the maximum bond dimension of MPS. 
With this flexibility, we can obtain the numerically precisely exact ground states 
of both the gapped and the gapless frustration-free Hamiltonians. 
It can also treat the long-range entangled spin liquid state up to the system size comparable to DMRG, 
and precisely identify all the topological sectors, which were difficult in DMRG. 
The advantage of injective TI MPS shall be to analyze the symmetry from a single tensor, 
while as an offset, 
the restriction onto the MPS tensor may be too strict to go beyond the area-law bound since it directly accesses the infinite-size limit. 
Because in our non-TI MPS, the condition imposed is minimal, the phase space shall be much more relaxed. 
}
In a similar context, the area law expected for MPS does not limit the representation of the finite-$N$ 
quantum many-body states even at finite temperatures following the volume law 
both in 1D and 2D\cite{Iwaki2021,Matthias2023}. 
\par
Exact solutions obtained on a large scale provide rigid theoretical starting points for exploring 
unknown quantum many-body phases. 
The spin liquid phase of a regular Heisenberg kagome antiferromagnet is discussed about 
three-coloring solutions\cite{Cepas2011,Changlani2018}, 
and the quantum scar states that get rid of thermalization can be studied by 
obtaining exact scar tower of states in the integer-spin 1D AKLT\cite{Moudgalya2018,Mark2020}. 
Our pragmatic protocol generating MPS systematically 
proves valuable for systems outside the particular regimes empirically known. 
Indeed, physically meaningful models for laboratory studies often necessitate maintaining the natural 
and standard form of the Hamiltonian, which makes the quantum state intriguing. 
Our approach achieves finding some particular points to be exactly solved 
in a class of frustration-free models without the need for additional numerical approximations or prior knowledge.

\appendix
\section{Frustration-free model as a quantum $k$-SAT problem}
\label{app:k-qsat}
Let us reformulate our scheme in connection with a $k$-satisfaction problem ($k$-SAT) of the classical complexity theory. 
In the $k$-SAT, there are $N$ Boolean variables $( x_1,\cdots, x_N)$ that are assigned true or false, 
and $M$ different constraints/clauses, each built as the OR ($\lor$) function of maximally $k$ variables. 
In the physics description, the variable $x_j$ can be represented by the spin variable pointing up 
($\sigma_i=+1$) and down ($\sigma_i=-1$) 
when $x_j$ is true and false, respectively. 
\par
A famous example is the 2-SAT formula for spin glass, where the Ising Edwards-Anderson Hamiltonian 
${\cal H}=\sum_{\langle i,j,\rangle} J_{ij}\sigma_i\sigma_j$ with randomly distributed $J_{ij}=\pm 1$ 
can be mapped to the sum of $k=2$ clause per bond, $(x_i\lor x_j)\land (\neg x_i\lor \neg x_j)$ for $J_{ij}=1$ and 
$(x_i \lor \neg x_j)\land (\neg x_i \lor x_j)$ for $J_{ij}=-1$ to have the satisfactory bonds with energy -1. 
The energy is counted as the number of satisfactory bonds, and the 2-SAT is 
whether we have the lowest energy state that satisfies all the $M$ clauses, which is known to be feasible; 
there is a critical value of $M/N$ above which almost all the answer becomes unsatisfactory\cite{Monasson1999}. 
If the model has three-body interactions $k=3$, whether or not the problem is satisfactory is NP-complete. 
Besides knowing that there are no satisfactory solutions, 
finding a spin glass state itself is an NP-complete problem. 
\par
The antiferromagnetic Ising model on a triangular lattice has a highly degenerate frustrated ground state. 
The 2-SAT problem of whether we find a solution that satisfies all bonds (frustration-free) or not is solved 
immediately and the answer is no, because one may define $3N$ bonds 
each with two clauses, $(x_i \lor x_j)\land (\neg x_i \lor \neg x_j)$, giving $M=6N$. 
However, finding a frustrated solution is reduced to a classical "frustration-free" problem. 
If we describe ${\cal H}$ not as a sum of bonds but as a sum of triangles, 
the energy of each antiferromagnetic triangle is the lowest when the system has either two-up-one-down or one-up-two-down spins, 
which is described as the combination of two clauses acting on three variables as, $(x_i \lor x_j \lor x_k) \land (\neg x_i \lor \neg x_j \lor\neg x_k)$. 
Whether we find a ground state that all the triangles satisfy the constraint as the 3-SAT problem\cite{Gosset2016} with $M=4N$. 
Although we already know the answer to be yes, there is no general polynomial algorithm to solve this NP-complete problem. 
The frustration-free problem we are considering is known as the quantum $k$-SAT. 
The $k$-SAT is included in the quantum $k$-SAT, 
namely even the quantum 2-SAT is much more difficult to solve than the classical counterpart. 
The range of clauses and $k$ that one could judge whether such quantum frustration-free solution will exist 
is being studied for some cases\cite{Movassagh2010,Sattath2016}, while the general strategy to 
have a solution is not being sufficiently understood. 
\par
We also notice that given a Hamiltonian, there are several ways to divide it into the form of Eq.(\ref{eq:ham}). 
Indeed, for a classical triangular lattice Ising antiferromagnet, setting $\hat h_l$ as each two-body term gives a frustrated UNSAT solution, 
whereas setting $\hat h_l$ as triangular unit consisting of three two-body interactions is a unfrustrated SAT one. 
In that respect, the problem of $k$-SAT or quantum $k$-SAT is being applied not to the Hamiltonian but to the 
way how the frustration-free form is constructed. 
\section{\revt{Orthogonality of the degenerate exact MPS states}}
\label{app:left-norm}
Here, we explain the details of step 4 (see Fig.~\ref{f5}(d)) in our MPS protocol. 
We truncate the bond dimensions of matrix of $D_n$ of $B_n^{i}$ 
to $\chi_n$ of $A_n^{i}$ one by one. 
Let the matrix be described as an assembly of column vectors of dimension $\chi_{n-1}$ as 
$B_n^{i}=(\bm b_{n,1}^{i},\cdots, \bm b_{n,D_n}^{i})$. 
Left-normalization imposed on these matrices indicates, 
\begin{eqnarray}
&& \sum_{i=1}^d B_{n}^{i \dagger} B_{n}^{i}= \hat I_{D_{n}}, \nonumber \\
&& \sum_{i=1}^d \bm b_{n,p}^{i\dagger }\bm b_{n,q}^{i} = \delta_{pq}, 
\label{eq:leftnorm}
\end{eqnarray}
where $\hat I_n$ is the unit matrix of dimension $n$. 
From among the $D_N$ bonds on the rightmost matrix $B_N$, 
choosing the $p$-th column vector $\bm b_{N,p}$, we find the $p$-th ground state $|\Psi_{N,p}^{\rm gs}\rangle$. 
Because of Eq.(\ref{eq:leftnorm}), for $p,q=1,\cdots, D_n$, 
\begin{equation}
\langle \Psi_{N,p}^{\rm gs}| \Psi_{N,q}^{\rm gs}\rangle 
= \bm b_{N,p}^\dagger \bm b_{N,q} =\delta_{pq}, 
\label{eq:psip_psiq}
\end{equation}
holds, namely, the degenerate ground states are orthogonal to each other. 
Unlike $B_n^{i}$, the $A_n^{i}$ obtained after the truncation 
for $n=1,\cdots,N-1$ are no longer shared by $D_N$ degenerate solutions. 

\section{\revt{Two other methods for obtaining PBC-MPS}}
\label{app:pbc-mps}
Here, in addition to \S.\ref{sec:pbcmps}, we explain two other methods, 
imposing projection or making use of the translational operator 
to obtain the PBC-MPS. 
\par
\begin{figure}
    \centering
    \includegraphics[width=8cm]{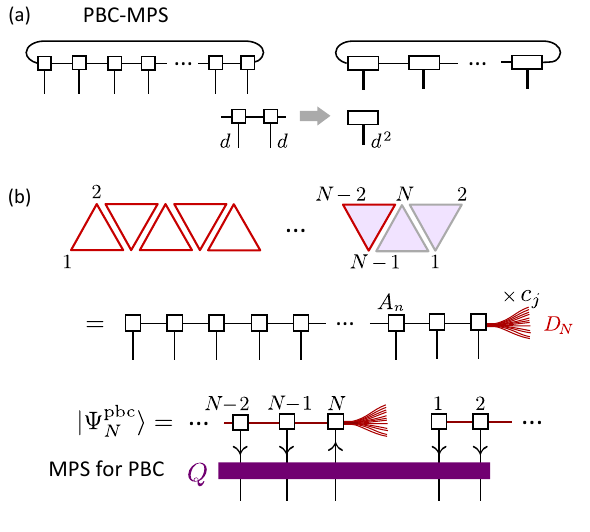}
    \caption{
    (a) Construction of 1D PBC-MPS. 
        The right panel is the $k=\pi$ PBC-MPS. 
    (b) Construction of MPS for PBC Hamiltonian that applies the same method for C-OBC MPS 
    by adding projections about the two triangles at the boundary.  }
    \label{f7}
\end{figure}

It is known that for a given OBC-MPS, one can construct a PBC-MPS with bond dimensions 
increased by a factor of $N$ \cite{Perezgarcia2007}; 
for each local degree of freedom, $i=1,\cdots d$, 
we combine a set of $\{A_n\}$ on the $(n,n+1)$ block as 
\begin{equation}
A^{{\rm pbc};i} = N^{-1/N} \left( \begin{array}{llllc}
 0 & A_1^{i} & 0 \cdots && 0 \\ 
 0 & 0 & A_2^{i} &  \\ 
 \vdots & & &\ddots \\
 0 & \cdots  &&& A_{N-1}^{i} \\
A_N^{i} & 0\cdots &&& 0\\
\end{array}\right), 
\label{eq:pbcmps-perez}
\end{equation}
which allows a newly obtained matrix of the form, 
\begin{eqnarray}
&&\sum_{\{i_n\}} {\rm tr} \big( \prod_{n=1}^N A^{{\rm pbc};i_n} \big) |i_N\cdots i_1 \rangle \nonumber \\
&&=\frac{1}{N}\sum_{j=0}^{N-1}\sum_{\{i_n\}}^d  
{\rm tr} \big( A^{i_1+j}_1 \cdots A^{i_N+j}_N \big) |i_N\cdots i_1 \rangle , 
\end{eqnarray}
which fulfills the translational invariance, and formally reduces to Eq.(\ref{eq:pbcmps1}). 
Such PBC-MPS is the eigenstate of the translation operator ${\cal T}$ and we can design it to have
${\cal T}|\Psi_{N}^{\rm pbc}\rangle=e^{ik} |\Psi_{N}^{\rm pbc}\rangle$. 
For example, when $k=\pi$, we combine two matrices as $(B_n^{i}\otimes B_{n+1}^{i})$ 
and prepare Eq.(\ref{eq:pbcmps-perez}) for $N/2$ blocks (see Fig.~\ref{f7}(a)). 
However, since the OBC-MPS (which includes C-OBC) in our case is degenerate, 
they naturally generate only part of the PBC-MPS, 
and does not guarantee the completeness of the solution. 
Another drawback is that the method requires more bond dimensions than necessary. 
Therefore, we propose three practical ways to obtain a full set of PBC-MPS. 
The most practically useful one is given in the main text. 
\par 
{\it Imposing projection.} 
This method does not differ much from the treatment of constructing C-OBC-MPS in \S.\ref{sec:1d-cobcmps}. 
We explain the difference using the zigzag ladder based on triangles in Fig.~\ref{f7}(b). 
The PBC Hamiltonian is realized from the C-OBC one by adding one triangle and connecting the edge sites. 
We follow step 1 and consecutively apply step 2 for $n=1,\cdots, N-2$ until we reach the last cluster. 
At the final step, we first add $B_{N-1}$ using the unit matrix, 
and then derive $B_{N}$ by simultaneously imposing three projection matrix $Q$ that operate 
on $(N-2, N-1, N)$, $(N-1, N, 1)$, and $(N, 1, 2)$ triangles. 
We then apply steps 4 and 5 and obtain $\tilde D_N$ degenerate solutions forming the columns of $\tilde A_N$. 
The two extra $Q$ required reduces the bond dimension to $\tilde D_N<D_N$. 
\par
{\it Translation operator.} \hspace{2mm}
The PBC solution is classified by an operator ${\cal T}$, 
that shifts the wave function by one lattice spacing along the 1D path of the MPS, 
where we assume that the $n=N$ site is put back to $n=1$. 
We first prepare a full set of OBC-MPS, $\{ |\Psi_{N,j}^{\rm gs}\rangle\}$, obtained in \S.\ref{sec:1d-cobcmps}. 
Then, we evaluate the matrix representation of $\langle \Psi_{N,j}^{\rm gs}|{\cal T} |\Psi_{N,k}^{\rm gs} \rangle$ 
of dimension $D_N$ and by diagonalizing it, 
the $\tilde D_N$ different PBC-MPS is obtained as its eigenstates. 
Namely, using the $l$-th eigen vector $(c_1^{(l)},\cdots, c_{D_N}^{(l)})$, 
the PBC-MPS takes the same form as Eq.(\ref{eq:pbcmps}), 
and the same process as Eq.(\ref{eq:pbclastmat}) is used to obtain $\tilde B_N^{i}$. 
The rest of the truncation process in step 4 is the same. 
\par
The three treatments all afford the PBC-MPS, while the numerical efficiency differs. 
For example, in constructing the PBC-MPS of the zigzag chain 
(see state {\bf E} in Fig.~\ref{f2}(a) in \S.\ref{sec:triangle}), 
we could reach up to $N=18$ for the method of imposing projection, 
while for diagonalizing $\mathcal{H}^{\rm bd}$ we can afford up to $N=40$, 
and find $D^{\rm PBC}_N=50$ degenerate PBC solutions out of $D_N=441$ C-OBC solutions. 
\begin{figure}
    \centering
    \includegraphics[width=9cm]{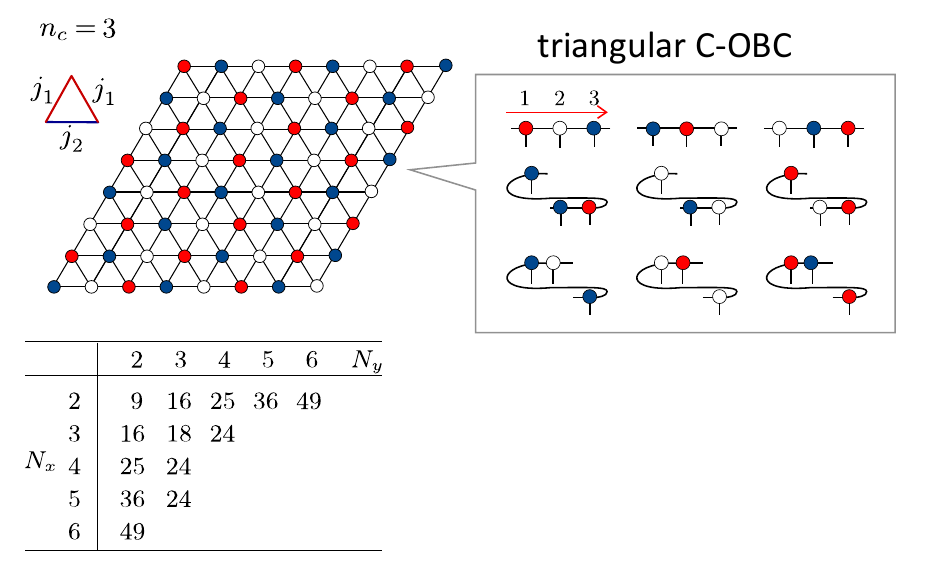}
    \caption{
      Results obtained for a parameter $\gamma=\sqrt{3}j$ with regular triangular geometry in Eq.(\ref{eq:hl_tri}) 
        corresponding to point {\bf E}. Unlike the zigzag chain, the parameter $\Gamma=\sqrt{3}J (=2\gamma)$ 
        also gives a regular triangle inside the lattice cluster. 
        Their ground state energy of $E=-3jN/4$
        In the lower panel, the degree of degeneracy of the ground state (bottom panel) is shown. 
        Example of the three sublattice ground states is shown whose MPS state consists of 9 tensors, 
        where the top three matrices are the constituents that do not cross the boundary. }
    \label{f9}
\end{figure}
%
\section{\revt{Three-colored product state on a triangular lattice}}
\label{app:tri-mps}
We construct MPS with $n_c=3, n_\cap=2$ for a triangular lattice of size $N=N_x\times N_y$ with C-OBC 
whose details are shown in Fig.~\ref{f9}. 
We consider a isosceles triangle cluster with coupling constants being $j_{ij}=j_1$ and $j_2$ and $\gamma_{ij}=\gamma_1$ and $\gamma_2$, 
whose Hamiltonian is 
$\hat h_l= \sum_{\langle m,n\rangle} j_{mn} \bm S_m \cdot \bm S_{n} + \gamma_{mm} (S^x_mS^y_{n}+ S^y_{m}S^x_{n})$. 
After we obtained the exact MPS ground states, 
we find that the three cases which have $M=2$ show the same series of $D_N$ as in Fig.~\ref{f9}. 
They are case {\bf D} with $j_2/j_1=-1/4$ and $\gamma=0$, 
case {\bf E} with $j_{mn}=j$, $\gamma_{mn}=\gamma$, 
and $\gamma=\sqrt{3}j$, and the $C_3$ triangular case, 
where we refer them in common to \S.\ref{sec:triangle} and Fig.~\ref{f2}. 
\par
It turned out that the states obtained for these triangular models 
share similarity with the exact ground states of the kagome lattice for 
the XXZ model at $J_z=-1/2$ \cite{Changlani2018} and the XYZ model 
($C_3$ line in Fig.~\ref{f2}(a)) \cite{Palle2021}, 
where they constructed the three-coloring exact solutions as product state. 
Indeed, in our case, the MPS tensors consist only of nine elements shown in Fig.~\ref{f8}(c), 
and the ground state are their product. 
\par 
To explicitly construct the three-sublattice product ground state in our framework, 
we introduce three different species of single-site states with indices $\gamma_s=a,b,c$ as 
\begin{align}
& |\gamma_s \rangle= \cos(\theta_{\gamma_s}/2) e^{-i\phi_{\gamma_s}/2} |1\rangle + \sin (\theta_{\gamma_s}/2) e^{i\phi_{\gamma_s}/2} |0\rangle, 
\label{eq:three-color}
\end{align}
which are parameterized by $\theta_{\gamma_s}$ and $\phi_{\gamma_s}$. 
For a unit triangle with its ground state $|\psi\rangle=|a\rangle|b\rangle|c\rangle$, 
these parameters are set to fulfill $\hat h_l |\psi\rangle =0$. 
Here, the cluster Hamiltonian is predefined for the XXZ and $C_3$ cases as 
$\hat h_l=|\xi_\Uparrow\rangle\langle \xi_\Uparrow|+|\xi_\Downarrow\rangle\langle \xi_\Downarrow|$. 
We then find the ground state $|\Psi^{\rm gs}\rangle=\prod_{i=1}^N \otimes |\gamma_s\rangle_{i\in \gamma_s}$. 
For the XXZ case, this form does not conserve the total $S^z$ so we need to project the obtained states 
to the local Hilbert space of each total-$S^z$ sector, which gives the exact ground state. 
\par
In the previously studied kagome lattice\cite{Changlani2018,Palle2021}, unlike the triangular lattice, 
there is a macroscopic number of configurations of tiling the three colors. 
However, the number of degeneracy can exceed the number of three colorings patterns 
for a sufficiently large-size lattice, which cannot be detected in their frameworks. 
Our method gives the unbiased exact degeneracy; 
for example, in a zigzag ladder ( $N_x$ or $N_y = 2$ case in Fig.~\ref{f9}), the number of three-colorings is restricted to 
maximally six, while we can still find a substantial degeneracy in the ground state, which cannot be 
captured by intuition.  

%
\section{Cluster Hamiltonian of a triangle with antisymmetric exchange term}
\label{app:triangle}
Here, we extend the parameter space of the choice of the cluster states 
from the ones shown in Eq.(\ref{eq:triangular_xi}). 
We include the parameters $\lambda,\mu,\nu$ to have the complex number of coefficients as 
\begin{eqnarray}
|\xi_\Uparrow\rangle &=& e^{i\lambda} \cos\alpha |000\rangle  
+ i \sin\alpha \big[e^{i\mu} \cos\beta|101\rangle 
\nonumber \\
&& \rule{3mm}{0mm}+\sin\beta \big(e^{i\nu} \cos\delta|110\rangle+\sin\delta|011\rangle \big) \big] ,
\nonumber \\
|\xi_\Downarrow\rangle &=& e^{-i\lambda}\cos\alpha |111\rangle 
- i \sin\alpha \big[e^{-i\mu} \cos\beta|010\rangle 
\nonumber \\
&& \rule{3mm}{0mm}+\sin\beta \big(e^{-i\nu}\cos\delta|001\rangle+\sin\delta|100\rangle\big) \big], 
\label{eq:triangular_xi-2}
\end{eqnarray}
Here, when we apply a local gauge transformation $U_z(\theta)=e^{-i\sigma_z\theta/2}$ 
to rotate the spin $xy$-axis by $\theta$ about the $z$-axis, 
which transforms the up and down spin states as $|0\rangle\rightarrow e^{-i\theta/2}|0\rangle$ 
and $|1\rangle\rightarrow e^{i\theta/2}|1\rangle$, 
we find 
\begin{equation}
\hat h_l(\alpha,\beta,\delta,\lambda,\mu,\nu) \rightarrow 
\hat h_l(\alpha,\beta,\delta,\lambda-2\theta,\mu,\nu), 
\end{equation}
which does not change the property of the Hamiltonian. 
Therefore, we set $\lambda=0$ and focus on the other two parameters. 
The resultant coupling constants include several terms that are not considered in the main text as 
\begin{equation}
\hat h_l=\sum_{\langle i,k\rangle=1,1',2} \sum_{\eta,\zeta=x,y,z} j_{ik}^{\eta\zeta} S_i^{\eta}S_k^{\zeta}
\end{equation}
with the exchange couplings for $\langle i,k\rangle=\langle 1,2\rangle$, 
$\langle 2,3\rangle$, and $\langle 1,3\rangle$ indexed as 1,1', and 2, respectively. 
They are given as
\begin{align}
j_1^{xx}&=j_1^{yy}= \sin^2\alpha\sin 2\beta \cos\delta\cos(\nu-\mu), \nonumber \\
j_1^{xy}&= \sin^2\alpha\sin 2\beta \cos\delta\cos(\nu-\mu)
+ \sin 2\alpha\sin \beta \sin\delta, \nonumber \\
j_1^{yx}&=-\sin^2\alpha\sin 2\beta \cos\delta\cos(\nu-\mu)
+ \sin 2\alpha\sin \beta \sin\delta, \nonumber \\
j_1^{zz} \;&= \cos^2\alpha + \sin^2\alpha (-\cos^2\beta+\sin^2\beta\sin^2\delta),\nonumber\\
j_{1'}^{xx}&= \sin^2\alpha\sin 2\beta \sin\delta\cos\mu - \sin 2\alpha\sin \beta \cos\delta\sin\nu , \nonumber \\
j_{1'}^{yy}&= \sin^2\alpha\sin 2\beta \sin\delta\cos\mu + \sin 2\alpha\sin \beta \cos\delta\sin\nu , \nonumber \\
j_{1'}^{xy}&= \sin^2\alpha\sin 2\beta \sin\delta\sin\mu
+ \sin 2\alpha\sin \beta \cos\delta\cos\nu, \nonumber \\
j_{1'}^{yx}&=-\sin^2\alpha\sin 2\beta \sin\delta\sin\mu
+ \sin 2\alpha\sin \beta \cos\delta\cos\nu, \nonumber \\
j_{1'}^{zz} \;&= \cos^2\alpha + \sin^2\alpha (-\cos^2\beta-\sin^2\beta\sin^2\delta),\nonumber\\
j_2^{xx}&= \sin^2\alpha\sin^2\beta \sin 2\delta\cos\nu 
 - \sin 2\alpha \cos\beta\sin\mu, \nonumber \\
j_2^{yy}&= \sin^2\alpha\sin^2\beta \sin 2\delta\cos\nu 
 + \sin 2\alpha \cos\beta\sin\mu, \nonumber \\
j_2^{xy}&= \sin^2\alpha\sin^2\beta \sin2\delta\sin\nu 
+ \sin 2\alpha\cos\beta \cos\mu, \nonumber \\
j_2^{yx}&=-\sin^2\alpha\sin^2\beta \sin2\delta\sin\nu 
+ \sin 2\alpha\cos\beta \cos\mu, \nonumber \\
j_2^z \;&= \cos^2\alpha +\sin^2\alpha \cos2\beta. 
\label{eq:tri_jparam}
\end{align}
In addition to the symmetric $\Gamma$-term that have equal $xy$ and $yx$ elements, 
the antisymmetric Dzyaloshinskii-Moriya interaction terms like $j^{xy}=-j^{yx}$ 
appear for this treatment. 
This will widely expand the model that provides the exact solution 
since the antisymmetric term ubiquitously appears in materials when the local bond inversion symmetry is lost.

\begin{figure}
    \centering
    \includegraphics[width=8cm]{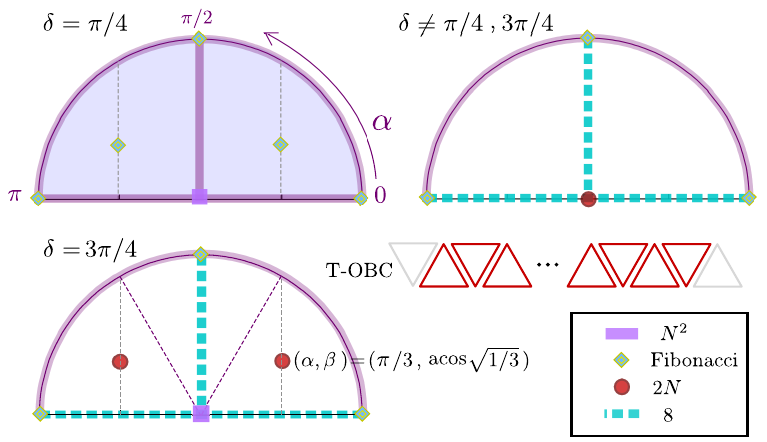}
    \caption{Solution space of the zigzag spin-1/2 chain based on a triangular unit 
   with the choice of $M=6$ in Eq.(\ref{eq:triangular_xi}), 
   where $\alpha$ is the tangent and $\beta$ denotes the vertical line as in Fig.~\ref{f2}(a). 
   For all cases, $\beta=[\pi/2:\pi]$ range has the same types of solution as the ones with $\beta=[0:\pi/2]$ 
   and is abbreviated. 
   The range without any symbols or lines has no exact solution. 
   $\delta=\pi/4, 3\pi/4$ and off these two are shown in different panels. 
}
    \label{f3}
\end{figure}
\section{\revt{Classification of the exact solutions}}
\label{sec:tr4}
We classify the solutions of the the zigzag chain with C-OBC in \S.\ref{sec:triangle} 
by the types of degeneracy in Fig.~\ref{f3}(a)-\ref{f3}(c) for $\delta=\pi/4, 3\pi/4$ and otherwise. 
We can add a unit triangle one by one to 
evaluate $D_N^{{\rm TOBC}}$ by rank$Q$ iteratively (see Ref.[42]). 
There are four types of degeneracy: 
$D_N$ increasing with order-$N^2$, with Fibonacci sequence, with $2N$, and constant $D_N=8$. 
Here, notice that by definition in Eq.(\ref{eq:triangular_xi}) 
once we take $\alpha=0,\pi$ the other two parameters $\beta,\delta$ do not make sense, 
and so as $\beta=0,\pi$ about $\delta$. 

{\it $N^2$-type degeneracy: }
The degeneracy increasing in a square appears in almost the whole region of the diagram at $\delta=\pi/4$  
except for a few points. 
In increasing the number of triangles linked and obtaining the form of $Q$, 
we find an iterative relationship, 
${\rm dim}(\cap_{l=1}^n V_l)/{\rm dim}(\cap_{l=1}^{n-1} V_l)=(n+4)(2(n+2))$ for even-$n$
and $(n+5)(2(n+3))$ for odd-$n$ with $n\leq N_c=N-2$ for C-OBC. 
Accordingly, we obtain the exact form, 
\begin{align}
    D_{N}^{{\rm TOBC}}=\left\{ 
    \begin{array}{ll}
    (N+2)^2/4 & ({\rm even}\;\;N)  \\ 
    (N+1)(N+3)/4 \;\;&  ({\rm odd}\;\;N). 
    \end{array}
    \right.
    \label{eq:DN_TOBC_square}
\end{align}
The degeneracy increasing in powers implies that the system is possibly 
gapless and is found to be a specific multicritical point, which is observed previously 
in some specific model parameters\cite{Batista2009,Batista2012,Changlani2018,Palle2021}. 
In the standard critical phase in quantum many-body systems, 
the ground state of a finite-size system is unique and has a finite-size gap that closes 
with $N^{-1}$ or $N^{-2}$\cite{Cardy1996}. 
However, when more than two phases meet at multicritical points, highly degenerate states can appear as ground states 
Fig.~\ref{f2} (c)-(III) {\bf B}, the multicritical point moves away from 
the highly symmetric model parameters when we vary $\alpha$ and $\beta$, 
and our solution can track them. 

{\it Fibonacci sequence:} 
Interestingly, degeneracy can sometimes form a Fibonacci sequence 
with increasing $N$ as $2,2,4,6,10,16 \cdots$, namely 
\begin{equation}
D_{N}^{{\rm TOBC}} =  2F_N,
\label{eq:DN_TOBC_Fibonacci}
\end{equation}
where $F_{N+2}=F_{N+1}+F_N$. 
These points appear at eight isolated points, $\alpha=0,\pi$, $(\alpha,\beta)=(\pi/2,0),(\pi/2,\pi)$, 
and $(\alpha,\beta)=(\pm\pi/3,{\rm acos}(\pm\sqrt{1/3}))$ with $\delta=\pi/4$. 
This degeneracy is an Ising-type\cite{Redner1981}, and increases exponentially with $N$.
\par
{\it $2N$-type:}
Solutions with degeneracy increasing linearly as $D_N=2N$ appear at 
the $\alpha=\beta=\pi/2$ and $\delta\ne\pi/4$ 
or  $\alpha=\pi/3,2\pi/3$, $\beta=\text{acos}(\sqrt{1/3})$ and $\delta=3\pi/4$. 
\par
{\it Constant-type:}
Along the lines of the $\alpha\beta$-plane at $\alpha=\pi/2$ and $\beta=\pi/2$ for $\delta\ne \pi/4$, 
the 8-fold degeneracy appears for $N\geq 4$. 

\begin{acknowledgements}
We thank Tomotoshi Nishino, Hosho Katsura, and Atsushi Iwaki for the informations. 
This work is supported by the "The Natural Laws of Extreme Universe" (No. JP21H05191) KAKENHI for Transformative Areas from JSPS of Japan, and JSPS KAKENHI Grants No. JP21K03440. 
\end{acknowledgements}
\bibliography{triangular_refs}

\end{document}